\documentclass{aa}

\usepackage[varg]{txfonts}
\usepackage{subfigure}
\usepackage{threeparttable}
\usepackage{natbib}
\usepackage{makecell}
\usepackage{multirow}
\usepackage{ulem}
\usepackage{graphicx}
\usepackage[colorlinks=true,linkcolor=blue,citecolor=blue,urlcolor=blue]{hyperref}
\bibpunct{(}{)}{;}{a}{}{,}

\begin{document}

   \title{Analysis of \textit{Gaia} Data Release 3 Parallax bias in the Galactic plane}

 
   \author{Ye Ding
          \inst{1,2}
          \and
          Shilong Liao\inst{1,2}\fnmsep\thanks{Corresponding author: Shilong Liao,shilongliao@shao.ac.cn}
          \and
          Qiqi Wu
          \inst{1,2}
          \and
          Zhaoxiang Qi
          \inst{1,2}
          \and
          Zhenghong Tang
          \inst{1,2}
         }

   \institute{Shanghai Astronomical Observatory, Chinese Academy of Sciences, Shanghai 200030, China
        \and
              University of Chinese Academy of Sciences, Beijing 100049, China
         }

   \date{}


 
  \abstract
   {The systematic errors are inevitable in \textit{Gaia} published astrometric data. Lindegren et al. (L21) proposed a global recipe to correct for the GEDR3 parallax zero point offset, which did not consider the Galactic plane. The applicability of their correction model to the Galactic plane remains uncertain. }
   {We attempt to have an independent investigation into the sample dependence of the L21 correction, and its applicability to the Galactic plane. }
   { We collect various samples, including quasars, binaries, and sources with parallaxes from other surveys or methods, to validate the L21 correction, especially in the Galactic plane.}
   { We conclude that the L21 correction exhibits sample dependence, and does not apply effectively to the Galactic plane. We present a new parallax bias correction applying to the Galactic plane, offering improvements over the existing L21 correction. The correction difference between L21 and this work can go up to 10 $\mu as$ within certain ranges of magnitude and colour. This work provides an additional recipe for users of \textit{Gaia} parallaxes, especially for sources located near the Galactic plane.}
   {}

   \maketitle
%

\section{Introduction}

The third data release from the European Space Agency mission \textit{Gaia} covers observations made between July 2014 and May 2017, during the first 34 months of the \textit{Gaia} mission \citep{2016A&A...595A...1G,2023A&A...674A...1G}.
It occurs in two stages: the early stage, \textit{Gaia} Early Data Release 3 (GEDR3), which provides updated astrometry and photometry, supplemented by radial velocities from the second data release (\textit{Gaia} Data Release 2, GDR2) \citep{2018A&A...616A...1G}; and the full \textit{Gaia} Data Release 3 (GDR3), which includes the same sources as GEDR3 along with new radial velocities, spectra, light curves, astrophysical parameters, and more. The astrometric data in GDR3 are identical to those in GEDR3. In GDR3, the typical uncertainties of parallaxes range from 0.03 to 1.3 mas for stars with 15 < G < 21 mag. 

As one of the most critical data provided by the \textit{Gaia} mission, high-precision, and high-accuracy parallaxes enable the derivation of reliable stellar distances. This has a significant impact on fields such as stellar formation and evolution, and the structure and dynamics of the Milky Way. However, systematic errors are inevitable in the published astrometric data because of imperfections in the instrument and data processing methods \citep{2021A&A...649A...2L}. 
Several studies have investigated the parallax zero point offset of GEDR3 using different tracers, including quasars ($\sim$ -21 $\mu$as, \citealt{2021A&A...649A...2L}; $\sim$ -24 $\mu$as, \citealt{2021PASP..133i4501L}), red giant stars ($\sim$ +22 $\mu$as, \citealt{2021AJ....161..214Z}), red clump stars ($\sim$ +26 $\mu$as, \citealt{2021ApJ...910L...5H}) and W Ursae Majoris (EW)-type eclipsing binary systems ($\sim$ -27 $\mu$as, \citealt{2021ApJ...911L..20R}). 
\citet{2021A&A...649A...5F}  compared the GEDR3 parallaxes with various external catalogues. With the use of  external reference catalogues of  open clusters (\citealt{2013A&A...558A..53K}; \citealt{2014A&A...564A..79D}), they derived an average zero point difference of 
-59 $\mu as$ and -91 $\mu as$, respectively.
\citeauthor{2021A&A...649A...4L} (\citeyear{2021A&A...649A...4L}, hereafter L21) proposed a recipe to correct the GEDR3 parallax zero point offset, which is a function of $G$-band magnitude, colour information and ecliptic latitude of the sources.
The correction functions are denoted as $Z_{5}(G, \nu_{eff}, \beta)$ and $Z_{6}(G, \hat{\nu}_{eff}, \beta)$ for five-parameter and six-parameter solutions, respectively, where $Z$ represents the parallax zero point offset.
The colour parameter used in L21 correction is the (photometric) effective wavenumber in $Z_{5}$, and the (astrometric) pseudocolour in $Z_{6}$. 
The L21 correction is based on three samples: quasars in \textit{Gaia}-CRF3 (hereafter GCRF3, \citealt{2022A&A...667A.148G}), stars in Large Magellanic Cloud, and binaries.
Thus, the applicability of the L21 correction to other samples not considered in their model remains unknown. Additionally, the samples they considered are nearly devoid of targets in the Galactic plane due to the high source densities in this region. Consequently, the applicability of their correction model to the Galactic plane also remains uncertain because of the paucity of sources in this area.

\begin{table*}
\caption{Statistics of the cross-matches of GDR3 to external catalogues.}           
\label{catalogs}      
\centering  
\begin{threeparttable}
\begin{tabular}{c c c c c c c c c}        
\hline\hline  
\multirow{2}{*}{Catalogue} & \multirow{2}{*}{Sources} & \multicolumn{2}{c}{Unique matches} & \multicolumn{2}{c}{Filtered sources} & \multicolumn{2}{c}{Retained sources}
& \multirow{2}{*}{Reference} \\

  &  & $\left | b\right | \leqslant {20}^{\circ}$ & $\left | b\right | > {20}^{\circ}$ 
  &  $\left | b\right | \leqslant {20}^{\circ}$ & $\left | b\right | > {20}^{\circ}$ 
  & $\left | b\right | \leqslant {20}^{\circ}$ & $\left | b\right | > {20}^{\circ}$  &  \\
  \hline
SDSS DR16Q & 750 414 & 2159 & 451 336 & 1851 & 386 161 & 53 & 4043 & \cite{2020ApJS..250....8L} \\
LAMOST DR10 & 83 376 & 2556 & 56 847 & 2005 & 54 743 & 155 & 609 & \cite{2023ApJS..265...25J} \\
LQAC-6 & 1 149 138 & 73 460 & 1 072 290 & 73 452 & 1 070 332 & - & 262 & {\cite{2024A&A...683A.112S}} \\
Quaia &1 295 502 &153 416 &1 142 086 & 153 256 & 1 141 270 & 39 425 & 128 280 & {\cite{2024ApJ...964...69S}} \\
GPQ & 100 153 & 100 104 & - & 100 087 & - & 20 468 & - & \cite{2021ApJS..254....6F,2022ApJS..261...32F}\\ 
CatNorth & 1 545 514 & 193 002 & 1 352 512 & 190 206 & 1 329 186 & 21 162 & 106 157 & \cite{2024ApJS..271...54F} \\
KiDS DR4 & 157 813 & - & 74 478 & - & 56 074 & - & 2237 & {\cite{2021A&A...649A..81N}} \\[2ex]

Wide Binary & 1 256 400 & 605 346 & 651 022 & 241 318 & 270 127 & 241 318 & 270 127 & \cite{2021MNRAS.506.2269E}\\[2ex]
           
VLBI & 108 & 75 & 27 & 70 & 25 & 70 & 25 &{\cite{2019ApJ...875..114X}} \\
HST & 111 & 46 & 56 & 34 & 41 & 34 & 41 & {\cite{2021A&A...654A..20G}}\\ 
Red Giant & 12 500 & 6821 & 7704 & 6456 & 7237 & 6456 & 7237 &{\cite{2023A&A...677A..21K}}\\
RR Lyrae & 401 & 77 & 324 & 76 & 324 & 76 & 324 & {\cite{2018MNRAS.481.1195M}}\\    
Red Clump & 137 448 & 96 813 & 37 666 & 91 929 & 36 508 & 91 929 & 36 508 & {\cite{2020ApJS..249...29H}}\\
EW Binary & 113 821 & 83 126 & 26 386 & 75 322 & 21 319 & 75 322 & 21 319 & {\cite{2021ApJ...911L..20R}} \\ 
RGB & 44 784 & 37 199 & 7410 & 36 514 & 7220 & 36 514 & 7220 &{\cite{2019AJ....158..147H}}\\
           
    \hline                                   
    \end{tabular}
\begin{tablenotes}
\footnotesize
    \item Notes. The external catalogues are listed in Col. 1. Column 2 shows the total number of sources in each catalogue. Columns 3-4 provide the number of unique matches in GDR3 for each catalogue. Columns 5-6 present the number of matched sources satisfying Eqs. (\ref{eq1})-(\ref{eq3}) for quasars, stars, and binaries, respectively. Columns 7-8 show the number of sources retained after filtering. Column 9 provides references for the catalogues. A dash (–) indicates that the number of sources is zero. 
\end{tablenotes} 
        
\end{threeparttable}
    
\end{table*}

The aim of this paper is to have an independent investigation into 
the sample dependence of the L21 correction, and its applicability to the Galactic plane. To achieve this, we have collected various samples within the Galactic plane. 
Quasars provide a direct estimate of the parallax bias, including data from the Large Sky Area Multi-Object Fibre Spectroscopic Telescope Data Release 10 (LAMOST DR10) \citep{2023ApJS..265...25J}, Sloan Digital Sky Survey Data Release 16 (SDSS DR16) \citep{2020ApJS..250....8L}, Sixth Release of the Large Quasar Astrometric Catalogue (LQAC-6) \citep{2024A&A...683A.112S}, the all-sky spectroscopic quasar catalogue - Quaia \citep{2024ApJ...964...69S},
the quasar candidate catalogue behind the Galactic plane (GPQ) \citep{2021ApJS..254....6F,2022ApJS..261...32F} and the improved \textit{Gaia} DR3 quasar candidate catalogue, CatNorth, built with data from \textit{Gaia}, Pan-STARRS1, and CatWISE2020 \citep{2024ApJS..271...54F}.
In addition, wide binaries \citep{2021MNRAS.506.2269E}, which are so widely separated that the two components never interact, have been considered to investigate the parallax bias.
What's more, parallaxes from other surveys or methods are required to investigate the parallax offset by comparing them with the \textit{Gaia} parallaxes, including (1) trigonometric parallaxes from the \textit{Hubble} Space Telescope (HST) \citep{2021A&A...654A..20G} and Very Long Baseline Interferometry (VLBI) astrometry \citep{2019ApJ...875..114X}, (2) RR Lyrae stars (RRLs) with the derived parallaxes from the recalibrated period-absolute magnitude-metallicity (PMZ) relations \citep{2018MNRAS.481.1195M}, (3) Red clump stars (RCs) with the derived parallaxes from recalibrated absolute magnitudes based on LAMOST data \citep{2020ApJS..249...29H}, (4) W Ursae Majoris (EW)-type eclipsing binary systems with the derived parallaxes from the recalibrated period–luminosity relation \citep{2021ApJ...911L..20R}, (5) Red giant stars (RGs) with the asteroseismic parallaxes \citep{2023A&A...677A..21K}, and (6) Red-giant branch  stars (RGBs) with the spectrophotometric parallaxes \citep{2019AJ....158..147H}.

The paper is structured as follows. Section \ref{sec2} details the sample selection process. In Sect. \ref{sec3}, we validate the L21 correction using the samples described in Sect. \ref{sec2}. Section \ref{sec4} presents a new estimate of the parallax correction coefficients applicable to the Galactic plane. In Sect. \ref{sec5}, we compare and validate two parallax bias corrections. The summary concludes the paper in Sect. \ref{sec6}.

\section{Sample selection\label{sec2}}
To investigate the sample dependence of the L21 correction and its applicability to the Galactic plane, we require extensive and high-quality samples, including QSOs, wide binaries, and sources with parallaxes from other surveys or methods. Table \ref{catalogs} lists the result of cross-matches with 15 external catalogues located at $\left | b\right | \leqslant {20}^{\circ}$ \footnote{We select $\left | b\right | \leqslant {20}^{\circ}$ rather than ${10}^{\circ}$ for two reasons: first, to broaden our sample coverage and encompass a larger portion of the Galactic plane, and second, to ensure an overlapping area with GCRF3 which was considered in L21.} and $\left | b\right | > {20}^{\circ}$ ($b$ represents Galactic latitude), respectively, followed by further astrometric filtering.

    \subsection{Cross-matching \label{sec2.1}}
    First, we cross-match the GDR3 astrometric catalogue with the external catalogues listed in Table \ref{catalogs}. The first seven entries in the table - SDSS DR16Q, LAMOST DR10, LQAC-6, Quaia, GPQ, CatNorth, and KiDS DR4 - are quasar catalogues. The eighth entry is the catalogue of binaries. Entries from the ninth to the last - VLBI, HST, RGs, RRLs, RCs, EWs,and RGBs - are sources with parallaxes from other surveys or methods. The astrometric information is used for the cross-matches. In cases where more than one \textit{Gaia} source is matched to a source in the external catalogue, and vice versa, these sources are removed. We only consider unique matches in the case where one \textit{Gaia} source is matched to one source in the external catalogue. Table \ref{catalogs} gives statistics of the cross-matches of GDR3 to external catalogues. Columns 3-4 list the number of unique matches located at $\left | b\right | \leqslant {20}^{\circ}$ and $\left | b\right | > {20}^{\circ}$ in GDR3, respectively. Further filtering, as described below, is then applied.

    \begin{figure*}
          \centering
          \subfigure{\includegraphics[width=0.33\linewidth]{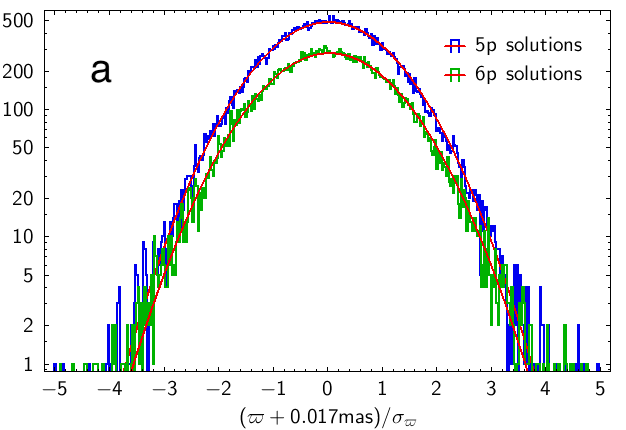}}
          \subfigure{\includegraphics[width=0.33\linewidth]{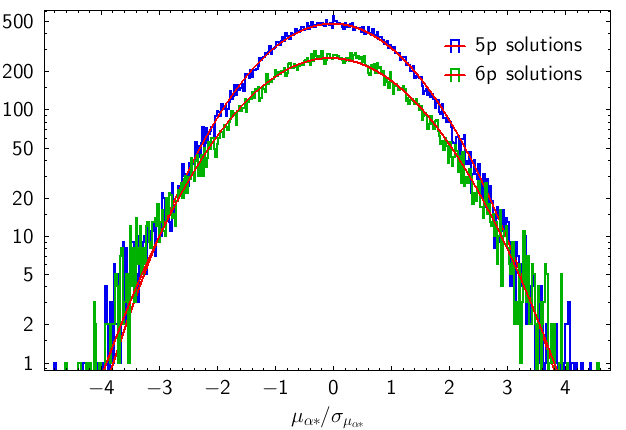}}
          \subfigure{\includegraphics[width=0.33\linewidth]{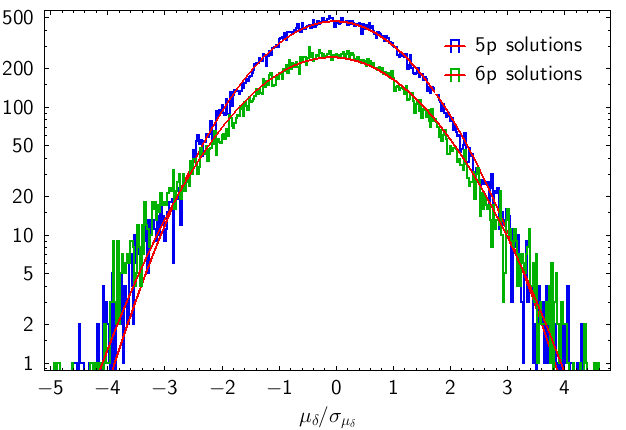}}
          \subfigure{\includegraphics[width=0.33\linewidth]{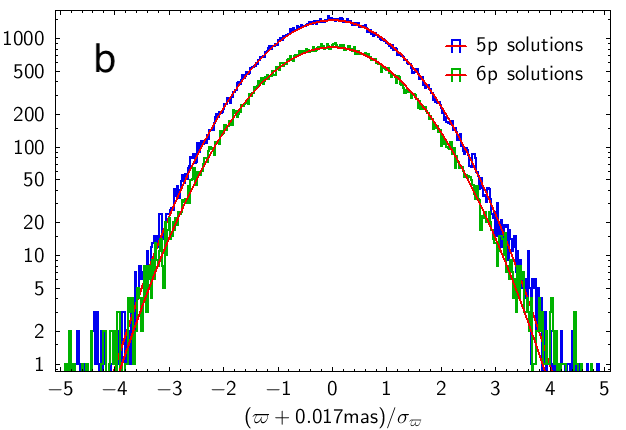}}
          \subfigure{\includegraphics[width=0.33\linewidth]{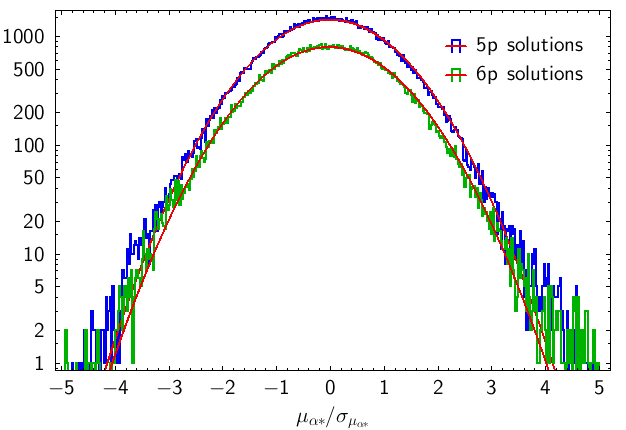}}
          \subfigure{\includegraphics[width=0.33\linewidth]{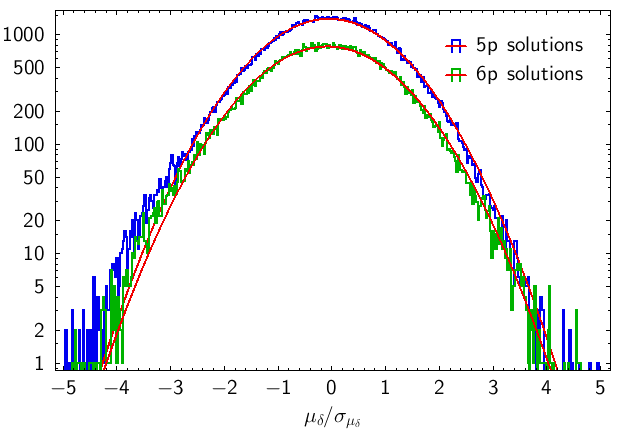}}
          \caption{Distributions of the normalised parallaxes and proper motion components for the Non-GCRF3 sample located at $\left | b\right | \leqslant {20}^{\circ}$ (a) and $\left | b\right | > {20}^{\circ}$ (b), respectively, with five-parameter (blue) and six-parameter (green) solutions. The turquoise curves show the corresponding best-fit Gaussian distributions.}
          \label{fig:quasar_normal}
\end{figure*}

    \subsubsection{Quasars}\label{sec2.1.1}
        Due to the high source densities in the Galactic plane, ensuring the cleanliness of the quasar sample is challenging.
        \cite{2020ApJS..250....8L} presented the final SDSS-IV quasar catalogue from Data Release 16, containing 750 414 sources in their quasar-only subcatalogue. We cross-match this catalogue with GDR3 by using the table `gaiadr3.dr2\_neighbourhood' in \textit{Gaia} archive, applying matching conditions of `angular\_distance < 100 mas' and `magnitude\_difference < 0.5 mag'. This results in 2159 and 451 336 unique matches located at $\left | b\right | \leqslant {20}^{\circ}$ and $\left | b\right | > {20}^{\circ}$, respectively.
        \cite{2023ApJS..265...25J} provided the LAMOST quasar catalogue from Data Releases 1 to 10 which contains 83 376 sources. We cross-match the LAMOST quasar catalogue with GDR3 at a radius of 2 arcsec, in which the numbers of unique matches located at $\left | b\right | \leqslant {20}^{\circ}$ and $\left | b\right | > {20}^{\circ}$ are 2556 and 56 847, respectively. 
        In addition, LQAC-6 presented by \cite{2024A&A...683A.112S} contains 1 149 138 sources with high quasar probability (>0.9). We cross-match LQAC-6 with GDR3 at a radius of 1 arcsec and get 1 145 750 unique matches. Among 1 145 750 matches, 73 460 and 1 072 290 sources are located at $\left | b\right | \leqslant {20}^{\circ}$ and $\left | b\right | > {20}^{\circ}$ respectively.
        \cite{2024ApJ...964...69S} presented the all-sky quasar catalogue, Quaia, that samples the largest comoving volume of any existing spectroscopic quasar sample. They improved the quasar candidate catalogue identified by the \textit{Gaia} mission with unWISE infrared data, applying cuts based on proper motions and colours. Their catalogue consists of 1 295 502 quasar candidates with G < 20.5, among which 153 416 and 1 142 086 sources are located at $\left | b\right | \leqslant {20}^{\circ}$ and $\left | b\right | > {20}^{\circ}$, respectively.

        In addition, we consider quasar catalogues achieved by machine learning. 
        \citet{2021ApJS..254....6F,2022ApJS..261...32F} provided a reliable GPQ candidate catalogue with 100 153 sources located at $\left | b\right | \leqslant {20}^{\circ}$ achieved through a transfer-learning method for quasar selection, with the purity of $\sim$ 90\% on the Simbad matches. Among them, 100 104 unique matches are considered.
        \cite{2024ApJS..271...54F} presented CatNorth, an improved GDR3 quasar candidate catalogue, built with data from \textit{Gaia}, Pan-STARRS1, and CatWISE2020. 
        Their catalogue contains 1 545 514 reliable quasar candidates from the parent GDR3 quasar candidate catalogue. Among them, 193 002 and 1 352 512 sources are located at $\left | b\right | \leqslant {20}^{\circ}$ and $\left | b\right | > {20}^{\circ}$, respectively.
        What's more, the quasar catalogue from the Kilo-Degree Survey (KiDS) Data Release 4 with a purity of 97\%, presented by \cite{2021A&A...649A..81N}, includes 157 813 sources with high quasar probability (p(QSOcand) > 0.9 for r < 22). We cross-match this catalogue with GDR3 at a radius of 2 arcsec, yielding 74 478 unique matches, all of which are located at $\left | b\right | > {20}^{\circ}$.
        
        
        

    \subsubsection{Wide binaries}\label{sec2.1.2}
        
        Similar to the approach in L21, we also consider binaries in our analysis. Physical binaries offer valuable insights into determining the parallax offset,  as the binary components have essentially near-identical parallaxes. The catalogue of wide binaries (WBs) of \cite{2021MNRAS.506.2269E} is listed in Table \ref{catalogs}. In its raw form, it contains 1.8 million candidate physical binaries. The sample is restricted to the subset of 1 256 400 pairs with >90\% probability of being bound (chance alignment <10\%). Among them, 605 346 and 651 022 pairs are located at $\left | b\right | \leqslant {20}^{\circ}$ and $\left | b\right | > {20}^{\circ}$, respectively, and further filtering will be described below.

    \subsubsection{Sources with parallaxes from other surveys or methods}\label{sec2.1.3}

        See Table \ref{catalogs}, initially, we consider independent trigonometric parallaxes.
        The VLBI astrometry is capable of measuring parallaxes with accuracies of $\sim$ 10 $\mu as$ \citep{2014ARA&A..52..339R}, which has an astrometric quality comparable to that of GDR3. We cross-match 108 VLBI sources of \cite{2019ApJ...875..114X} with GDR3 at a radius of 100 mas, retaining 75 and 27 unique matches located at $\left | b\right | \leqslant {20}^{\circ}$ and $\left | b\right | > {20}^{\circ}$, respectively. In cases where one VLBI source has multiple reference parallaxes, we retain the one with higher precision (e.g., VY CMa, HD 283447, CoKu HP Tau G2, V1096 Tau, V1023 Tau, HD 283572, IRAS 20126+4104, HD 226868). 
        \cite{2021A&A...654A..20G} compiled a list of 111 objects with the independent trigonometric parallax data from HST, considering the results determined by using the Fine Guidance Sensor (FGS) and the Wide Field Camera 3 (WFC3) from several studies (\citealt{2017PASP..129a2001B}; \citealt{2020PASP..132e4201V}; \citealt{2018ApJ...853...55B}; \citealt{2014ApJ...785..161R,2018ApJ...855..136R}; \citealt{2016ApJ...825...11C}).
        We use the results of \cite{2021A&A...654A..20G} which provided GDR3 identifiers in their table. Of the 111 objects, only one object is not listed in GDR3 (Polaris A), and 8 objects have no parallax listed in GDR3. Among the left 102 objects, 46 and 56 unique matches are located at $\left | b\right | \leqslant {20}^{\circ}$ and $\left | b\right | > {20}^{\circ}$, respectively. 
        
        In addition to independent trigonometric parallaxes, derived parallaxes from different methods are also considered. First, independent distance measurements are available from asteroseismology of red-giant stars. \cite{2023A&A...677A..21K} determined distances to nearly 12 500 RGs observed by Kepler, K2, and TESS. After cross-matching with GDR3 at a radius of 2 arcsec, 6821 and 7704 unique matches are located at $\left | b\right | \leqslant {20}^{\circ}$ and $\left | b\right | > {20}^{\circ}$, respectively. 
        
        Additionally, derived parallaxes making use of standard candles, distance indicators, and spectrophotometry are considered as well, including RRLs (\citealt{2018MNRAS.481.1195M}), RCs (\citealt{2020ApJS..249...29H}), EWs (\citealt{2021ApJ...911L..20R}), and RGBs (\citealt{2019AJ....158..147H}). 
        First, RRLs are radially pulsating stars used as standard candles for distance measurement. \cite{2018MNRAS.481.1195M} provided derived parallaxes for 401 RRLs based on recalibrated period-absolute magnitude-metallicity (PMZ) relations and absolute magnitude-metallicity relations. Among them, 77 and 324 RRLs are located at $\left | b\right | \leqslant {20}^{\circ}$ and $\left | b\right | > {20}^{\circ}$, respectively. Primary RCs are metal-rich low-mass stars (typically smaller than 2 $M_{\odot}$) with stable luminosities, making them reliable distance indicators. 
        \cite{2020ApJS..249...29H} presented a sample of 137 448 RCs with derived parallaxes. Using over 10 000 primary RCs with GDR2 parallaxes, they recalibrated the $K_{S}$ absolute magnitudes based on LAMOST data, deriving distances for all RCs with this recalibration. The purity and completeness of their sample are generally higher than 80\%. After cross-matching with GDR3 at a radius of 2 arcsec, 96 813 and 37 666 unique matches are located at $\left | b\right | \leqslant {20}^{\circ}$ and $\left | b\right | > {20}^{\circ}$, respectively. 
        EWs can also serve as distance indicators due to their well-defined period-luminosity relationship. \cite{2021ApJ...911L..20R} presented a catalogue of 113 821 EWs with derived parallaxes from the recalibrated period-luminosity relation. Among them, 83 126 and 26 386 EWs are located at $\left | b\right | \leqslant {20}^{\circ}$ and $\left | b\right | > {20}^{\circ}$ in GDR3, respectively. 
        \cite{2019AJ....158..147H} determined spectrophotometric distances to 44 784 RGBs predicted with a data-driven model combining spectroscopy from APOGEE DR14 and photometric information from 2MASS, \textit{Gaia}, and WISE. After cross-matching with GDR3 at a radius of 2 arcsec, 37 199 and 7410 unique matches are located at $\left | b\right | \leqslant {20}^{\circ}$ and $\left | b\right | > {20}^{\circ}$, respectively.
        The distance indicator calibrations of RRLs, RCs, EWs, and RGBs, as mentioned above, indirectly depend on \textit{Gaia} DR2 or DR3 parallaxes, making it potentially circular to use them for verifying GDR3 parallaxes. Therefore, it is important to note that these sources can not be considered independent checks. However, they remain valuable for investigating parallax zero-point differences as a function of colour and sky position.

    \subsection{Quality filtering\label{sec2.2}}
    
    \subsubsection{Quasars}\label{sec2.2.1}
        For all quasar catalogues in Table \ref{catalogs}, we apply the following criteria, which are adopted from the GCRF3 quasar selection \citep{2022A&A...667A.148G}.
    
        \begin{align}\label{eq1}
        \left\{     
        	\begin{aligned}
        	(i) \quad &astrometric\_params\_solved = 31 \quad or \quad 95, \\
                (ii) \quad &\left | (\varpi +0.017 mas )\right |/ {\sigma }_{\varpi }<5,  \\
                (iii) \quad &{{\mathrm{X}}_{\mu}}^{2} \equiv  \begin{bmatrix}\mu_{\alpha*} & \mu_{\delta} \end{bmatrix}{Cov(\mu )}^{-1}\begin{bmatrix} \mu_{\alpha*} \\ \mu_{\delta} \end{bmatrix} < 25
        	\end{aligned}
        \right.
        \end{align}
        Criterion (i) selects the sources that have five- or six-parameter astrometric solutions; Criterion (ii) takes the global parallax zero point offset of EDR3 (-17 $\mu as$) \citep{2021A&A...649A...4L} into account; Criterion (iii) is that the proper motion components $\mu_{\alpha*}$ and $\mu_{\delta}$ should be small compared to their uncertainties $\sigma_{\mu \alpha*}$ and $\sigma_{\mu \delta}$, where ${Cov(\mu )}^{-1}$ is the covariance matrix of proper motion. 
        The numbers of unique matches that survived the filtering by Eq. (\ref{eq1}) are provided in Cols. 5-6 of Table \ref{catalogs}. This filtering resulted in the removal of 14\%, 22\%, 0.01\%, 0.1\%, 0.02\%, and 1\% of sources for SDSS DR16Q, LAMOST DR10, LQAC-6, Quaia, GPQ, and CatNorth within the Galactic plane, respectively, and 14\%, 4\%, 0.2\%, 0.07\%, 2\% and 25\% of sources for SDSS DR16Q, LAMOST DR10, LQAC-6, Quaia, CatNorth, and KiDS DR4 outside the Galactic plane, respectively.

    \subsubsection{Wide binaries}\label{sec2.2.2}

        It can be assumed that the components of binaries have similar true parallaxes, although their magnitudes and colours may be very different. Thus, the parallax differences between the components of binaries should be as close to zero as possible. For the WBs, we applied the following criteria.

        \begin{align}\label{eq2}
        \left\{     
        	\begin{aligned}
                (i)\quad &astrometric\_params\_solved1,2 = 31 \quad or\quad 95 \\
                (ii)\quad &\left|(\varpi_{1}-\varpi_{2}) \right| / \sqrt{{{\sigma }_{\varpi_{1}}}^{2}+{{\sigma }_{\varpi_{2}}}^{2}} < 5 \\
                (iii)\quad & 16.1 <= G_{2} < 21.0 
        	\end{aligned}
        \right.
        \end{align}
        where `1' and `2' denote the components with the brighter and fainter G magnitudes, respectively. Criterion (i) selects WBs whose components both have five- or six-parameter astrometric solutions. Criterion (ii) selects WBs whose parallax differences between two components fall within five times the formal uncertainty. Criterion (iii) selects WBs whose fainter components can be corrected using the quasar-derived recipe (see Sect. \ref{sec4}). The numbers of unique matches that survived the filtering by Eq. (\ref{eq2}) are listed in Cols. 5-6 of Table \ref{catalogs}, resulting in the removal of 60\% and 59\% of pairs within and outside the Galactic plane, respectively.

    \subsubsection{Sources with parallaxes from other surveys or methods}\label{sec2.2.3}
    
        First, for sources with parallaxes from other surveys or methods in Table \ref{catalogs}, we applied the criteria (i) in Eq. (\ref{eq1}) to select the sources that have five- or six-parameter astrometric solutions. We assume that the parallax offsets ($\Delta \varpi$) of these sources are given by the difference between GDR3 parallax ($\varpi_{G}$) and other parallax determinations ($\varpi_{O}$). Assuming that errors are Gaussian as given in GDR3, we use the following criteria to select sources, 
        \begin{align}\label{eq3}
             \left |  (\varpi_{G} - \varpi_{O}) \right | / \sqrt{{{\sigma }_{\varpi_{G}}}^{2}+{{\sigma }_{\varpi_{O}}}^{2}} <5
        \end{align}
        where the parallax difference of the source should be zero within five times the formal uncertainty. The numbers of unique matches of sources that survived the filtering by Eq. (\ref{eq1}-i) and Eq. (\ref{eq3}) are given in Cols. 5-6 of Table \ref{catalogs}, removing 7\%, 26\%, 1\%, 7\%, 9\%, 5\%, and 2\% of sources within the Galactic plane, and 7\%, 27\%, 0, 3\%, 19\%, 6\%, and 3\% of sources outside the Galactic plane, for VLBI, HST, RRLs, RCs, EWs, RGs, and RGBs, respectively.

    \subsection{Final selection \label{sec2.3}}

    The numbers of the retained sources are given in Cols. 7-8 of Table \ref{catalogs}. 
    For WBs, all filtered sources are retained, with 63\% having five-parameter solutions and 37\% having six-parameter solutions. For sources with parallaxes from other surveys or methods, all filtered sources are also retained, with 98\% having five-parameter solutions and 2\% having six-parameter solutions.
    For all quasar catalogues, we first exclude overlapping sources in GCRF3 to obtain the Non-GCRF3 sources. Next, duplicate Non-GCRF3 sources among quasar catalogues are removed. The final union of the Non-GCRF3 sample contains 81 263 sources located at $\left | b\right | \leqslant {20}^{\circ}$ and 241 588 sources at $\left | b\right | > {20}^{\circ}$. Among the Non-GCRF3 sources within the Galactic plane, 51 546 (63\%) have five-parameter solutions and 29 717 (37\%) have six-parameter solutions in GDR3. Outside the Galactic plane, 153 923 (64\%) have five-parameter solutions and 87 665 (36\%) have six-parameter solutions in GDR3.

    To estimate the level of stellar contamination in Non-GCRF3, we analysed histograms of the normalised parallaxes $\varpi / {\sigma_{\varpi} }$ and normalised proper motion components $\mu_{\alpha*} / \sigma_{\mu_{\alpha*}}$ and $\mu_{\delta} / \sigma_{\mu_{\delta}}$. Ignoring small systematic and random offsets, the normalised quantities should follow Gaussian distributions with mean values close to zero and standard deviations only slightly larger than one for a clean sample of QSOs, as described in \cite{2022A&A...667A.148G} (see Sect. 2.3). Conversely, if their distributions deviate significantly from the expected Gaussian, it indicates that the sample is contaminated by stellar objects with non-zero parallaxes and/or proper motions.
    %
    Figure \ref{fig:quasar_normal} shows the distributions of the normalised parallaxes and proper motion components for the Non-GCRF3 sample located at $\left | b\right | \leqslant {20}^{\circ}$ and $\left | b\right | > {20}^{\circ}$, respectively, with five-parameter and six-parameter solutions.
    For the Non-GCRF3 sample located at $\left | b\right | \leqslant {20}^{\circ}$, the standard deviations of the three normalised quantities are 1.058, 1.075, and 1.098 for the five-parameter solutions and 1.071, 1.158, and 1.210 for the six-parameter solutions. For the Non-GCRF3 sample located at $\left | b\right | > {20}^{\circ}$, the standard deviations are 1.042, 1.087, and 1.107 for the five-parameter solutions and 1.053, 1.105, and 1.127 for the six-parameter solutions.
    In comparison, the standard deviations of the normalised quantities in GCRF3 are 1.052, 1.055, and 1.063 for the five-parameter solutions and 1.073, 1.099, and 1.116 for the six-parameter solutions \citep{2022A&A...667A.148G}. 
    The slightly larger standard deviations in our Non-GCRF3 sample indicate minor differences, but overall, the distributions remain very symmetric and close to Gaussian, suggesting a low level of stellar contamination.
    
\begin{figure}
          \centering
          \subfigure{\includegraphics[width=0.9\linewidth]{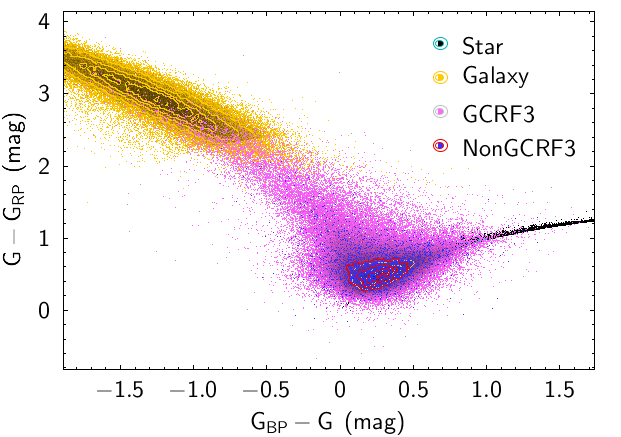}}
          \caption{Colour–colour diagram for Non-GCRF3 (blue), GCRF3 (pink), and galaxies (yellow) as well as stars (black) in the Galactic plane ($\left | b\right | \leqslant {20}^{\circ}$). 
          } 
          \label{fig:quasar_color_color}
    \end{figure}

\begin{figure}
          \centering
          \subfigure{\includegraphics[width=0.9\linewidth]{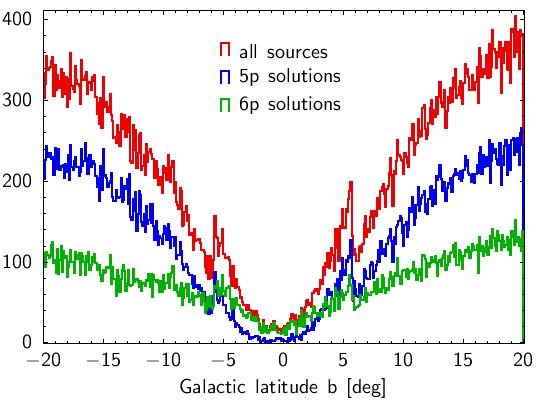}}
          \caption{Distribution of the Non-GCRF3 sample located at $\left | b\right | \leqslant {20}^{\circ}$ in Galactic latitude. The distribution is shown for the whole sample (red) and separately for the sources with five-parameter (blue) and six-parameter (green) astrometric solutions.}
          \label{fig:quasar_b}
    \end{figure}

    Due to the high density of stellar sources in the Galactic plane, we also plot the colour-colour diagram for the Non-GCRF3 sample located at $\left | b\right | \leqslant {20}^{\circ}$, compared with galaxies, stars, and GCRF3 located at $\left | b\right | \leqslant {20}^{\circ}$, see Fig. \ref{fig:quasar_color_color}. For the Non-GCRF3 sample, we sample 2000 quasars to better visualise the overlapping structures with GCRF3. 
    The selection of galaxies and stars used in the analysis are detailed in Appendix \ref{appendix:a}. The overlap between the contours of the Non-GCRF3 and GCRF3  samples indicates the reliability of our sample.
    Figure \ref{fig:quasar_b} displays the distribution of the Non-GCRF3 sample located at $\left | b\right | \leqslant {20}^{\circ}$ in Galactic latitude.
    In this paper, we focus on sources with five-parameter solutions, as most of our samples have five-parameter solutions.

        \begin{figure}
              \centering
              \subfigure{\includegraphics[width=0.85\linewidth]{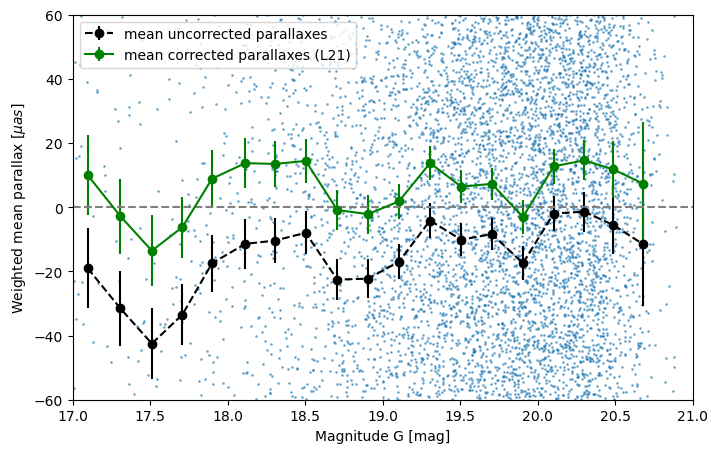}}
              \subfigure{\includegraphics[width=0.85\linewidth]{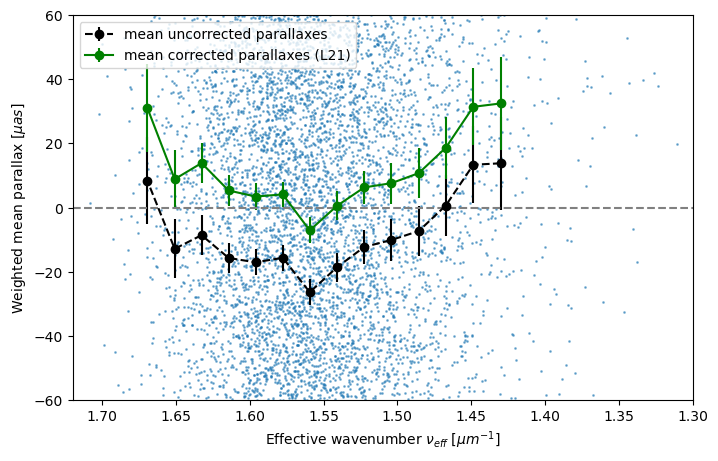}}
              \subfigure{\includegraphics[width=0.85\linewidth]{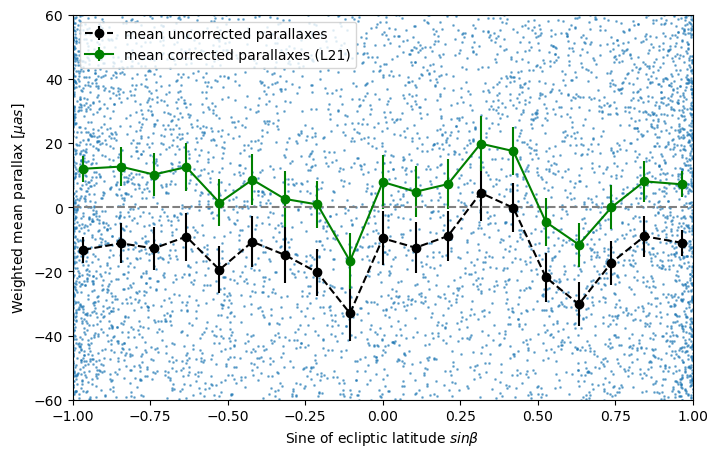}}
              \caption{Parallaxes for 51 546 quasars with five-parameter solutions located at $\left | b\right | \leqslant {20}^{\circ}$ in the Non-GCRF3 sample. }
              \label{fig:test_l21_qso_l20}
        \end{figure}
        
        \begin{figure}
              \centering
              \subfigure{\includegraphics[width=0.85\linewidth]{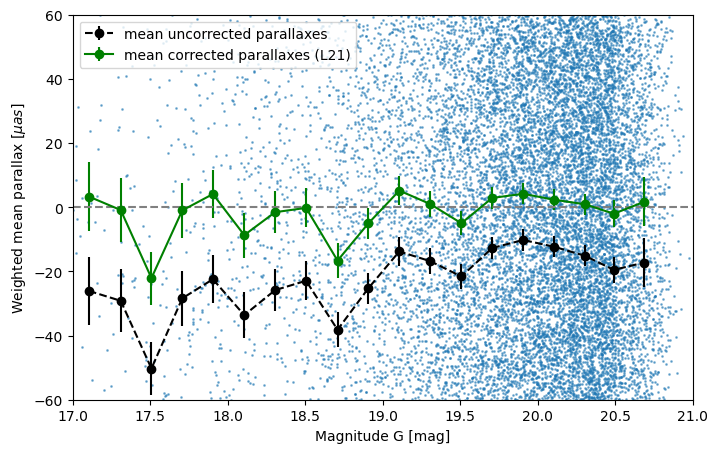}}
              \subfigure{\includegraphics[width=0.85\linewidth]{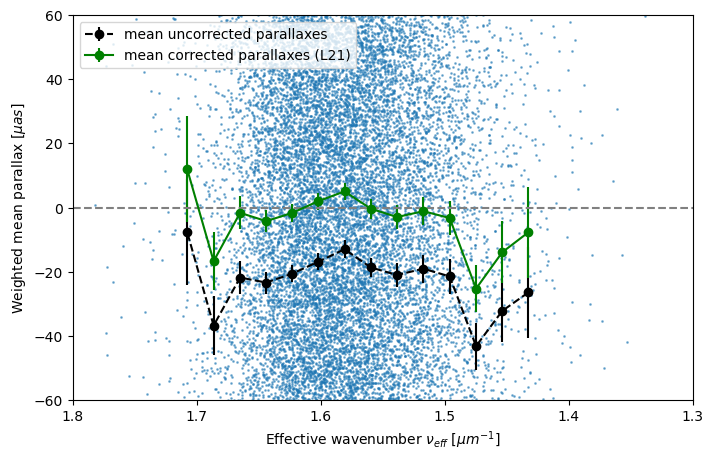}}
              \subfigure{\includegraphics[width=0.85\linewidth]{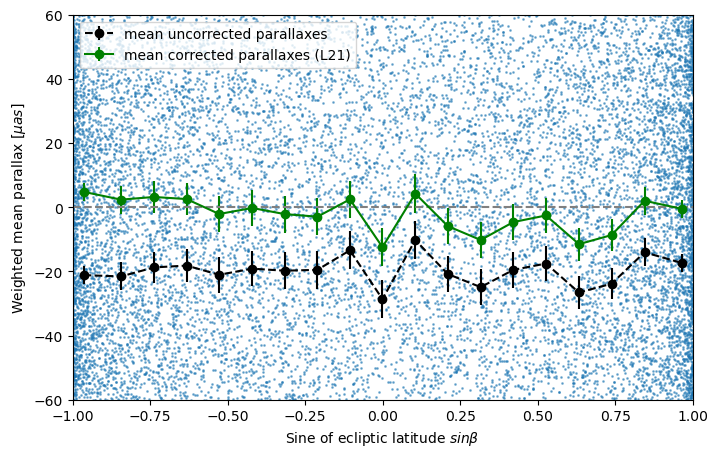}}
              \caption{Parallaxes for 153 923 quasars with five-parameter solutions  located at $\left | b\right | > {20}^{\circ}$ in the Non-GCRF3 sample.}
              \label{fig:test_l21_qso_g20}
        \end{figure}

\section{Validation of the L21 correction\label{sec3}}
     In this section, we assess the validity of the L21 correction by dividing the sky into two regions: the Galactic plane region ($\left | b\right | \leqslant {20}^{\circ}$) and the off-Galactic plane region ($\left | b\right | > {20}^{\circ}$).
     We use QSOs, WBs, and sources with parallaxes from other surveys or methods listed in Table \ref{catalogs} to evaluate the L21 correction.

    \subsection{Quasar verification}\label{sec3.1}
        First, we use quasars to assess the validity of the L21 correction. Since the L21 correction is primarily based on the GCRF3 sample, we apply it to the Non-GCRF3 sample to examine its sample dependence.
        
        Figures \ref{fig:test_l21_qso_l20} and \ref{fig:test_l21_qso_g20} display the results for the five-parameter solutions of the Non-GCRF3 sample located at $\left | b\right | \leqslant {20}^{\circ}$ and $\left | b\right | > {20}^{\circ}$, respectively. The figures are divided according to magnitude, colour, and ecliptic latitude. Blue dots show the individual values plotted versus magnitude, effective wavenumber, and sine of ecliptic latitude. Black circles show mean values of the uncorrected parallaxes ($\varpi$) in bins of magnitude etc.; green circles show mean values of the corrected parallaxes ($\varpi-Z_{5}$) applying the L21 correction.
        In Fig. \ref{fig:test_l21_qso_l20}, the corrected parallaxes (green) are not centred on zero, showing a zero-point deviation of approximately 20 $\mu as$. It indicates that the L21 correction does not apply to the Galactic plane effectively. In contrast, Fig. \ref{fig:test_l21_qso_g20} shows that the L21 correction is more effective for sources located at $\left | b\right | > {20}^{\circ}$, particularly in regions of high source densities.

    \subsection{Binary verification}\label{sec3.2}
    
        It can be assumed that the components of binaries have similar true parallaxes, although their magnitudes and colours may be very different. We compare the corrected parallax differences (applying the L21 correction) of WBs with the zero point to check the validity of the L21 correction. 

\begin{figure}
              \centering
              \subfigure{\includegraphics[width=0.85\linewidth]{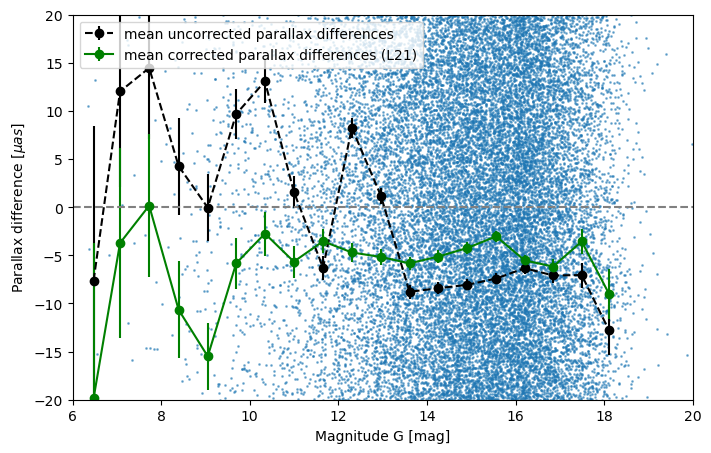}}
              \subfigure{\includegraphics[width=0.85\linewidth]{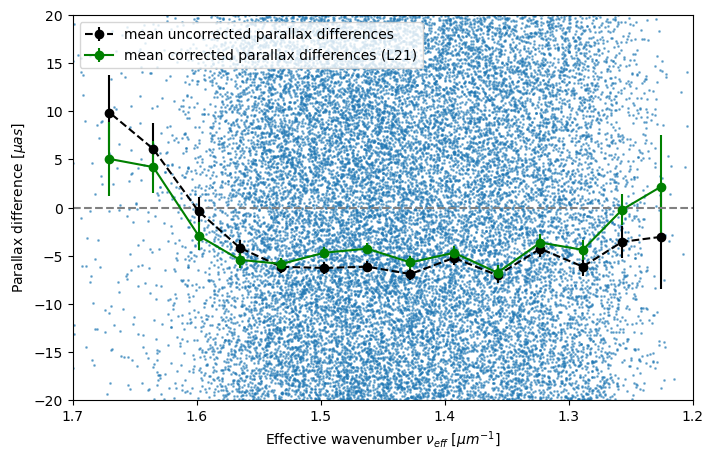}}
              \subfigure{\includegraphics[width=0.85\linewidth]{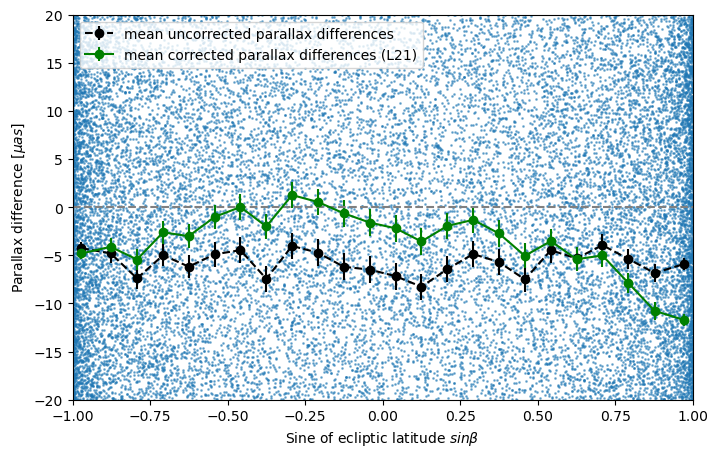}}
              \caption{Parallax differences of 148 234 binaries located at $\left | b\right | \leqslant {20}^{\circ}$ with five-parameter solutions binned by magnitude, colour, and ecliptic latitude of the bright components of binaries. }
              \label{fig:test_l21_wb_l20}
        \end{figure}

\begin{figure}
              \centering
              \subfigure{\includegraphics[width=0.85\linewidth]{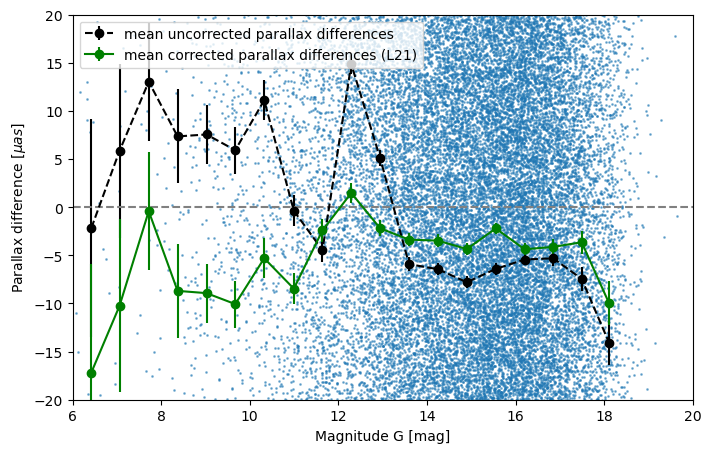}}
              \subfigure{\includegraphics[width=0.85\linewidth]{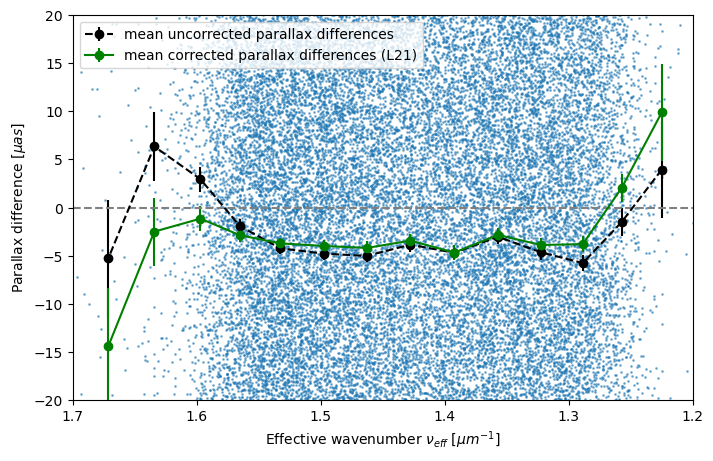}}
              \subfigure{\includegraphics[width=0.85\linewidth]{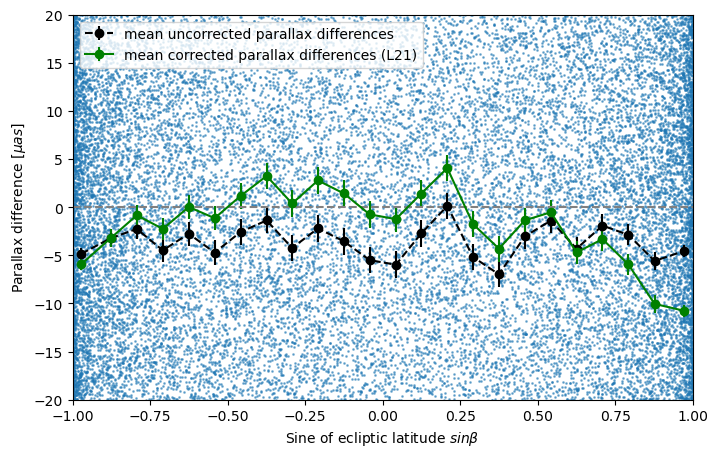}}
              \caption{Parallax differences of 172 931 binaries located at $\left | b\right | > {20}^{\circ}$ with five-parameter solutions binned by magnitude, colour, and ecliptic latitude of the bright components of binaries.}
              \label{fig:test_l21_wb_g20}
        \end{figure}

        Figures \ref{fig:test_l21_wb_l20} and \ref{fig:test_l21_wb_g20} show the parallax differences for the five-parameter solutions of WBs located at $\left | b\right | \leqslant {20}^{\circ}$ and $\left | b\right | > {20}^{\circ}$, respectively, divided according to magnitude, colour, and ecliptic latitude. The horizontal axes represent the properties of the bright components of the binaries.
         Blue dots show the individual values plotted versus magnitude, effective wavenumber, and sine of ecliptic latitude. Black circles show mean values of the uncorrected parallax differences ($\varpi_{1}-\varpi_{2}$) in bins of magnitude etc.; green circles show mean values of the corrected parallax differences applying the L21 correction ($(\varpi_{1}-Z_{5_{1}})-(\varpi_{2}-Z_{5_{2}})$).
        After applying the L21 correction, the mean values of parallax differences (green) of WBs are not centred on zero, regardless of whether WBs are located at $\left | b\right | \leqslant {20}^{\circ}$ or $\left | b\right | > {20}^{\circ}$. We conclude that the L21 correction exhibits sample dependence and does not apply effectively to these WBs. 
        The parallax differences as a function of magnitude and colour differences are shown in Appendix \ref{appendix:parallax diff}. As expected, these differences approach zero for small magnitude and colour differences for both the corrected and uncorrected versions.

    \subsection{Other verification}\label{sec3.3}
     In addition to quasar and binary verifications, we also compare independent parallaxes from other surveys or methods with the GDR3 parallaxes to assess the validity of the L21 correction, including VLBI, HST, and RG sources listed in Table \ref{catalogs}. 
     Table \ref{table:test_l21_star} gives the results for five-parameter solutions of these sources located at $\left | b\right | \leqslant {20}^{\circ}$ and $\left | b\right | > {20}^{\circ}$, respectively. 
     In the Galactic plane ($\left | b\right | \leqslant {20}^{\circ}$), the L21 correction results in overcorrection for VLBI sources. Outside the Galactic plane ($\left | b\right | > {20}^{\circ}$), the L21 correction results in overcorrection for VLBI, HST, and RG sources. For VLBI and HST sources, the sample sizes are too small to assess the L21 correction firmly. Based on these results, we conclude that the L21 correction exhibits sample dependence.
                
     Although RRLs, RCs, EWs, and RGBs cannot be considered independent checks due to their indirect dependence on \textit{Gaia} parallaxes, we still present the results for these sources in Table \ref{table:test_l21_star}. Additionally, we plot the parallax zero-point differences as a function of colour and ecliptic latitude for sources with derived parallaxes (RGs, RRLs, RCs, EWs, and RGBs) in Figure \ref{fig:valid_star}. Details see Appendix \ref{appendix:star_diff}. Overall, the parallax zero-point differences show distinct behaviours within and outside the Galactic plane, suggesting the necessity of this work.


    \begin{table*}
    \caption{Parallax corrections for sources with parallaxes with five-parameter solutions.}           
    \label{table:test_l21_star}      
    \centering
    \begin{threeparttable}
         \begin{tabular}{c c c c c c c}     
            \hline\hline         
            \multirow{2}*{Catalogue} & \multicolumn{3}{c}{$\left | b\right | \leqslant {20}^{\circ}$} & \multicolumn{3}{c}{$\left | b\right | > {20}^{\circ}$} \\
                           & \makecell{$\Delta \varpi$ (uncorrected parallax \\ offset) ($\mu as$) }
                           & \makecell{L21 correction \\ ($\mu as$) }  & N      
                           & \makecell{$\Delta \varpi$ (uncorrected parallax \\ offset) ($\mu as$) }
                           & \makecell{L21 correction \\ ($\mu as$) }  & N          \\
           \hline
           VLBI & -8.22 & +21.5   & 32  & +13.06 & +33.26  & 9 \\
           
           HST & -22.96  & +1.43 & 16 & +111.93 & +137.58  & 24 \\
           
           Red Giant & -19.15  & +2.21  & 6300 & -15.94 & +8.20  & 6859 \\
           
           RR Lyrae & -31.76  & -1.92 & 74 & -30.33 & +1.55  & 303 \\
           
           Red Clump & -25.98  & +4.64  & 90 452 & -20.65 & +7.51  & 35 484 \\

           EW binary & -19.30  & +14.87  & 75 296 & -15.69 & +18.61  & 21 317 \\

           RGB & -19.21  & +9.97   & 27426 & -25.86 & +0.43  & 6989 \\
        \hline                                   
        \end{tabular}
        \begin{tablenotes}
        \footnotesize
            \item Notes. Column 1 gives the name of catalogue. Columns 2-4 and 5-7 present the results for sources located at $\left | b\right | \leqslant {20}^{\circ}$ and $\left | b\right | > {20}^{\circ}$, respectively. Columns 2 and 5 show the weighted mean of the offset between the observed GDR3 parallaxes and parallaxes from other surveys or methods. Columns 3 and 6 give the weighted mean of the L21 correction. Columns 4 and 7 indicate the number of sources with five-parameter solutions.
        \end{tablenotes} 
    \end{threeparttable}
    \end{table*}

\section{Estimate of the new correction coefficients\label{sec4}}

In this section, we fit the parallax bias function for sources with five-parameter solutions and get the new coefficients applying to the Galactic plane, primarily based on parameterised function $Z(G, \nu_{eff},\beta)$ defined in L21 Appendix A. 
The generic function is
\begin{align}\label{eq:bias_function}
        Z(G, \nu_{eff},\beta) = \displaystyle\sum_{j} \displaystyle\sum_{k} q_{jk}(G) \, c_{j}(\nu_{eff}) \, b_{k}(\beta)
\end{align}
where the functions 
\begin{align}\label{eq:bias_function}
        q_{jk}(G) = \displaystyle\sum_{i} z_{ijk} \, g_{i}(G), \; j = 0...4, \; k = 0...2
\end{align}
are piecewise linear in G. Among the functions, $g_{i}(G)$ are the basis functions in magnitude, $c_{j}(\nu_{eff})$ are the basis functions in colour, and $b_{k}(\beta)$ are the basis functions in ecliptic latitude. The coefficients $z_{ijk}$ are the free parameters used to fit $Z$ to the given data. For more details on the basis functions, see L21 Appendix A. 

The primary methods for estimating the new correction are: (i) using quasars to directly estimate the parallax bias for $G$ $\gtrsim$ 16, and (ii) using binaries to map differential variations for bright stars.

\subsection{Quasars\label{sec4.1}}
First, the Non-GCRF3 sample with five-parameter solutions located at $\left | b\right | \leqslant {20}^{\circ}$ is used to estimate the new correction coefficients. 
For completeness, we also consider GCRF3 in the estimation. The GCRF3 sample located at $\left | b\right | \leqslant {20}^{\circ}$ contains 139 907 sources with five-parameter solutions. The combined sample includes 191 286 sources with five-parameter solutions and wavenumbers in the range 1.24 to 1.72 $\mu m^{-1}$. The magnitudes range from $G \simeq$ 13.8 to 21.0; only 99 are brighter than $G$ = 16.1.

\begin{figure}

      \subfigure{\includegraphics[width=0.85\linewidth]{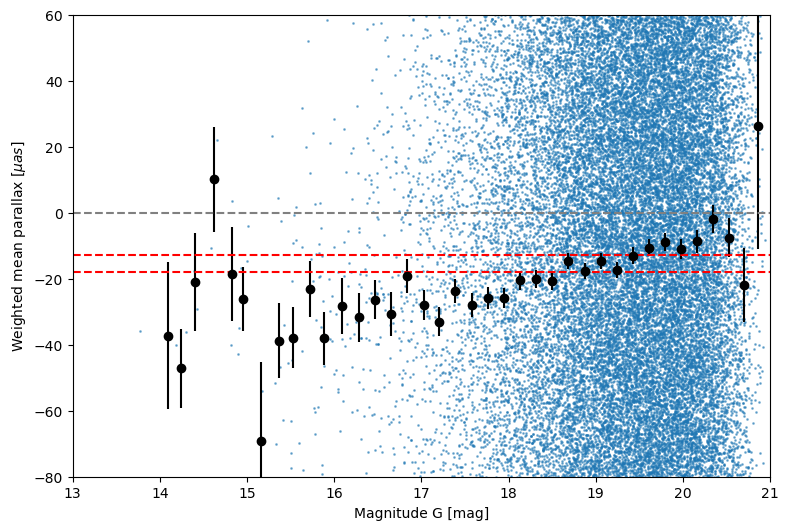}}
      \subfigure{\includegraphics[width=0.85\linewidth]{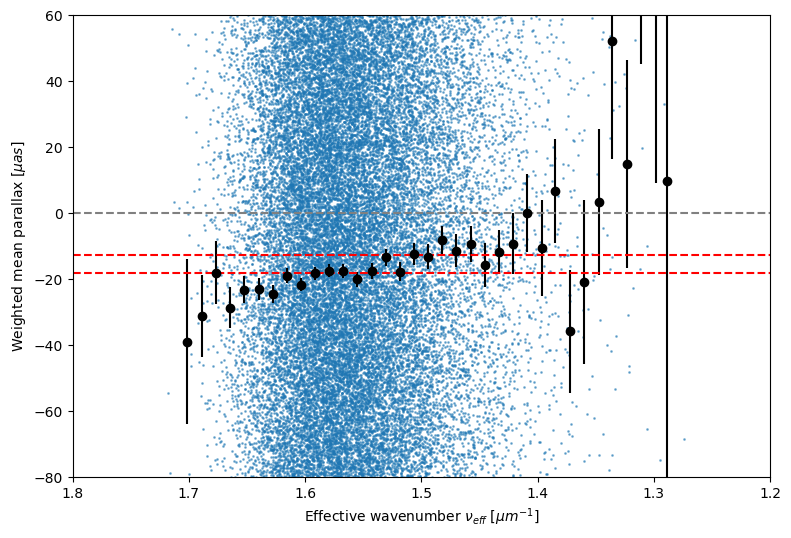}}
      \subfigure{\includegraphics[width=0.85\linewidth]{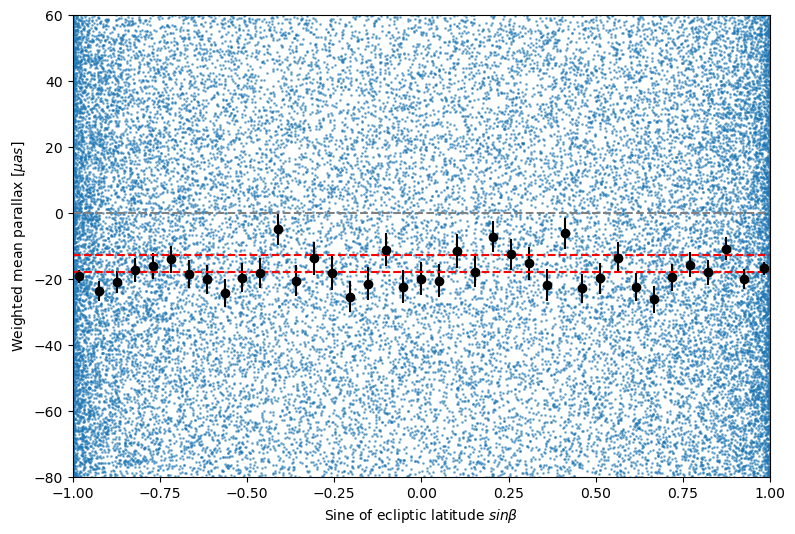}}
      \caption{Mean parallax of quasars binned by magnitude, effective wavenumber, and sine of ecliptic latitude. The two red lines indicate the weighted mean value (-18 $\mu as$) and median (-12.7 $\mu as$) of the full sample.}
      
      \label{fig:quasar}
\end{figure}

Figure \ref{fig:quasar} shows the weighed mean parallax plotted against $G$, $\nu_{eff}$, $\beta$. The blue points show the parallaxes of individual sources. In each bin, the black square is the mean parallax in DR3 weighted by $\sigma_{\varpi}^{-2}$, with error bars indicating the estimated standard deviation of the weighted mean. The main trends are as follows.

      \begin{itemize}
      \item Overall, a negative parallax bias is observed: the weighted mean parallax of the sample is approximately -18 $\mu as$, with a median of about -12.7 $\mu as$. These values are represented by the red lines in Fig. \ref{fig:quasar}, and they are slightly smaller than those reported in L21 (-21 and -17 $\mu as$).
      \item As a function of $G$, there is a non-linear variation when $G \lesssim$ 16, followed by an approximately linear increase from $G \backsimeq$ 16 to 20.
      \item The variation as a function of $\nu_{eff}$ shows an approximately linear trend within the well-populated range of colours.
      \item As a function of $\beta$, there is a non-linear variation.
      \end{itemize}

The trends described above can be well approximated by the parameterised function $Z(G, \nu_{eff},\beta)$ in Eq. (\ref{eq:bias_function}). However, considering the limited support in $G$ and $\nu_{eff}$, we need to add some constraints. We only use i = 8...12 for the basis function in magnitude, $g_{i}(G)$, for the scarcity of quasars in the range $G <$ 16.1. Additionally, for the basis function in colour, $c_{j}(\nu_{eff})$, we use j = 0 and 1 for wavenumbers of the sample in the range 1.24 to 1.72 $\mu m^{-1}$. The quasar sample contains few sources redder than $\nu_{eff} \simeq$ 1.4 $\mu m^{-1}$ or bluer than 1.7 $\mu m^{-1}$, and therefore cannot be used to estimate $q_{jk}$ for j = 2, 3, and 4. 

\begin{table}
\caption{Coefficients for the function fitted to the quasar parallaxes.}           
\label{table:quasar_fit}    
\centering
\begin{threeparttable}
\setlength{\tabcolsep}{4pt} 
    \begin{tabular}{c c c c c c c}     
    \hline\hline                 
      $G$ & $q_{00}$ & $q_{01}$ & $q_{02}$ & $q_{10}$ & $q_{11}$ & $q_{12}$  \\ 
      \hline
      16.1 & -28.46 & +16.35  & -39.53 & +47.80  & -159.86 & +136.25   \\
       
      17.5 & -20.79  & -6.36  & -2.21 & -55.98 & +41.07 & -20.35   \\
       
      19.0 & -11.49  & +4.50  & -1.85 & -45.32  & -26.16 & +5.02   \\
       
      20.0 & -5.03  & -0.69  & -5.61 & -57.69  & +49.24 & +76.35  \\
       
      21.0 & -2.29  & -2.01  & +51.04 & -112.23  & +123.50 & -261.33   \\
      \hline                                   
    \end{tabular}

    \begin{tablenotes}[flushleft]
    \footnotesize
        \item Notes. The table gives $q_{jk}(G)$ at the values of $G$ in the first column. For other values of $G$, linear interpolation should be used. Units are: $\mu as$ (for $q_{0k}$) and $\mu as$ $\mu m$ ($q_{1k}$).
    \end{tablenotes} 
\end{threeparttable}
\end{table}

Our estimate of $Z_{5}(G, \nu_{eff},\beta)$ from the quasar sample is given by the coefficients $q_{jk}$ in Table \ref{table:quasar_fit}. It is primarily valid within the subspace populated by quasars. Most importantly, this does not include sources that are brighter than $G$ $\simeq$ 16 or redder than $\nu_{eff}$ $\simeq$ 1.48 $\mu m^{-1}$. To extend $Z_{5}$ in these directions, we resort to the differential method using physical pairs.


\begin{figure}
      \centering
      \subfigure{\includegraphics[width=0.85\linewidth]{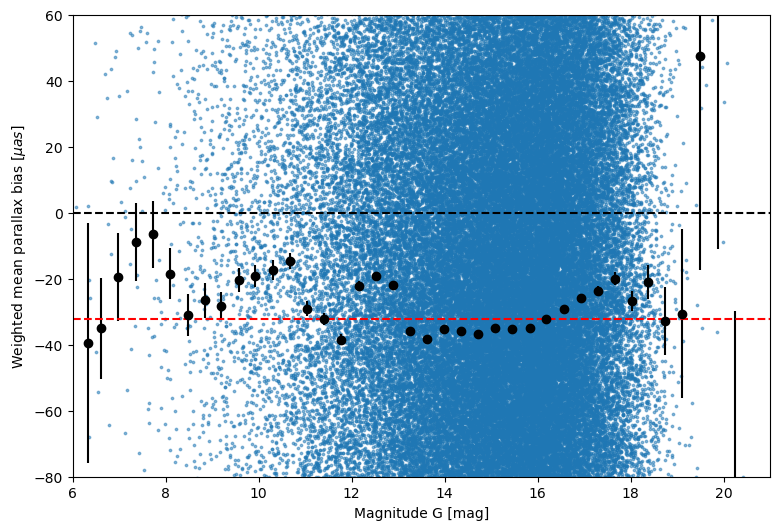}}
      \subfigure{\includegraphics[width=0.85\linewidth]{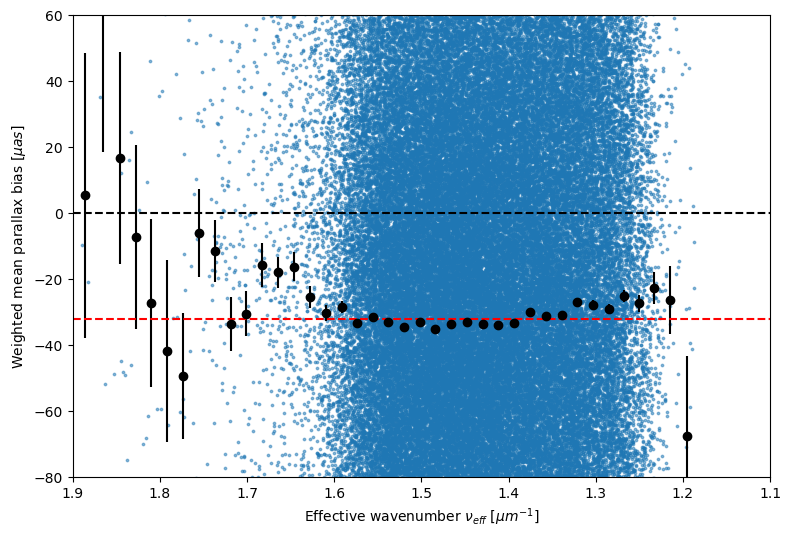}}
      \subfigure{\includegraphics[width=0.85\linewidth]{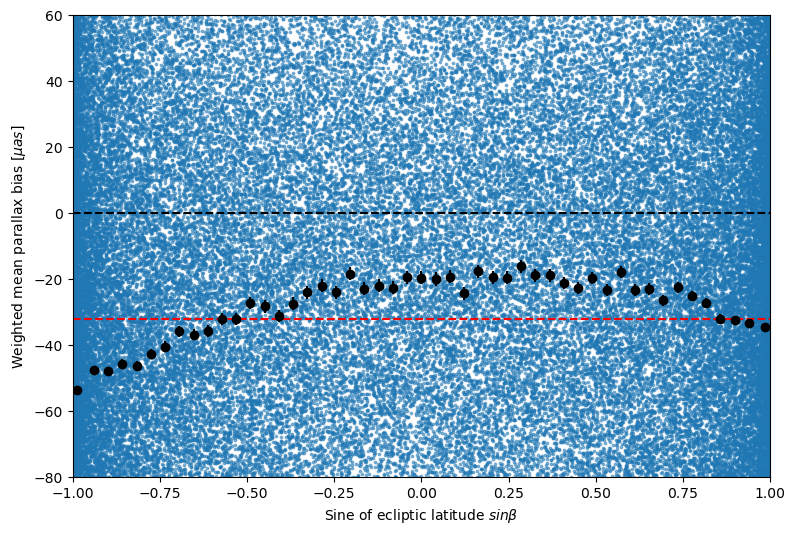}}
      \caption{Weighed mean parallax bias of the bright components of binaries binned by magnitude, effective wavenumber, and sine of ecliptic latitude. The two overlapping red lines indicate the weighted mean (-31.3 $\mu as$) and median (-29.1 $\mu as$) of the full sample.}
      \label{fig:binaries}
\end{figure}

\subsection{Wide binaries\label{sec4.2}}

    Having mapped $Z_{5}(G, \nu_{eff},\beta)$ for $G$ > 16.1 by the mean of quasars, we extend our analysis to brighter sources.
    Similar to the methods in L21, we also use binaries, in which it can be assumed that the components share similar true parallaxes. Using the results from the previous section to 'anchor' the parallax bias of the fainter component of binaries, we can then  estimate the bias of the brighter component, as shown in Eq. (\ref{eq:wb_bias}).
    \begin{align}\label{eq:wb_bias}
        Z_{5}(G_{1}, \nu_{eff1},\beta) = \varpi_{1} - \varpi_{2} +  Z_{5}(G_{2}, \nu_{eff2},\beta)
    \end{align}
    where `1' and `2' denote the components with the brighter and fainter G magnitude of binaries respectively, and $Z_{5}(G_{2}, \nu_{eff2},\beta)$ represents the parallax bias of the faint component, with $Z_{5}$ defined by the coefficients in Table \ref{table:quasar_fit}. The bias-corrected parallax of the faint component is computed as $\varpi_{2} - Z_{5}(G_{2}, \nu_{eff2}, \beta)$, where $\varpi_{2}$ and $\nu_{eff2}$ are the published GDR3 parallax and effective wavenumber of the faint component. $Z_{5}(G_{1}, \nu_{eff1},\beta)$ is denoted as the parallax bias estimate of the brighter component.
    We use WBs listed in Table \ref{catalogs} with five-parameter solutions and $G$ in the range of 6.0 to 21.0, located at $\left | b\right | \leqslant {20}^{\circ}$, to estimate the new correction coefficients. 

    Figure \ref{fig:binaries} shows the parallax bias of the bright components plotted against $G$, $\nu_{eff}$, $\beta$. The blue points show the parallax differences of individual sources. In each bin, the black square is the weighed mean parallax difference in DR3 weighted by $\sigma_{\varpi}^{-2}$, with error bars indicating the estimated standard deviation of the weighted mean. The main trends are as follows.

      \begin{itemize}
      \item In general, there is a negative parallax bias: the weighted mean parallax of the sample is approximately -31.3 $\mu as$, with a median of about -29.1 $\mu as$, as indicated by the two overlapping red lines in Fig. \ref{fig:binaries}. 
      \item As a function of $G$, there is a non-linear variation, except an approximately linear increase from $G \backsimeq$ 16 to 18.
      \item As a function of $\nu_{eff}$, the variation is approximately linear when $\nu_{eff}$ $\gtrsim$ 1.5 and $\lesssim$ 1.7 $\mu m^{-1}$, with a slight curvature when $\nu_{eff}$ $\gtrsim$ 1.2 and $\lesssim$ 1.5 $\mu m^{-1}$. The sources are very few when $\nu_{eff}$ $\gtrsim$ 1.7 $\mu m^{-1}$.
      \item As a function of $\beta$, the variation can be described by a quadratic polynomial in $\sin{\beta}$.
      \end{itemize}

    The investigated magnitude interval overlaps with the range in Table \ref{table:quasar_fit} for $G$ $\simeq$ 16.1 to 19, providing a consistency check and potentially improved estimate of the parallax bias in an interval poorly covered by quasars. Considering the trends in Fig. \ref{fig:binaries}, We use i = 0...10 for the magnitude basis function, $g_{i}(G)$, and j = 0,1,2 for the colour basis function, $c_{j}(\nu_{eff})$. 
    The curvature with colour in the mid-red region ($\nu_{eff}$ = 1.24 to 1.48 $\mu m^{-1}$) is fully described by the non-interaction term $q_{20}$, that is $q_{21}$ = $q_{22}$ = 0. 
    A severe limitation is that the scarcity of bright sources with $\nu_{eff}$ < 1.2 or > 1.7 $\mu m^{-1}$ makes it practically impossible to determine $q_{30}$ and $q_{40}$. The resulting fit is given in Table \ref{table:binaries_fit}.

\begin{table*}
    \caption{Coefficients for the function as estimated from wide binaries.}     
    \label{table:binaries_fit}      
    \centering
    \begin{threeparttable}
        \begin{tabular}{c c c c c c c c}     
        \hline\hline                 
       $G$ & $q_{00}$ & $q_{01}$ & $q_{02}$ & $q_{10}$ & $q_{11}$ & $q_{12}$ & $q_{20}$ \\ 
           \hline
           6.0 & -17.82  & +18.56 & -31.07 & -53.63 & +149.36 & -298.61 & -4877.35   \\
           
           10.8 & -13.26  & +13.52 & -30.77 & +20.86 & -10.82 & +53.55 & -2412.08   \\
           
           11.2 & -23.51  & +9.61 & -32.61 & -73.44 & +88.29 & -14.61 & +13501.35 \\
           
           11.8 & -31.06  & +16.19 & -27.02 & -44.66 & -27.31 & +44.23 & -4053.40 \\
           
           12.2 & -9.10  & +13.07 & -30.86 & -95.25 & +18.22 & +97.42 & +5590.58   \\
    
           12.9 & -11.70  & +17.77 & -34.43 & -38.31 & -83.65 & +16.80 & +56.79   \\
    
           13.1 & -30.48  & +15.50 & -31.92 & -7.78 & -67.85 & +16.60 & -272.16   \\
    
           15.9 & -33.38  & +6.62 & -20.93 & -82.04 & -48.82 & +43.17 & -2002.14   \\
    
           16.1 & -35.26  & +3.76 & -18.14 & -115.44 & -50.20 & +49.47 & -1960.32  \\
    
           17.5 & -44.82  & -14.78 & -2.64 & -237.22 & -31.98 & -41.61 & -1560.55   \\
    
           19.0 & -39.90  & +38.81 & -18.97 & +66.18 & +37.32 & -143.85 & +3162.85   \\
          \hline                                   
        \end{tabular}

        \begin{tablenotes}
        \footnotesize
            \item Notes. The table gives $q_{jk}(G)$ at the values of $G$ in the first column. For other values of $G$, linear interpolation should be used. Units are: $\mu as$ (for $q_{0k}$) and $\mu as$ $\mu m$ ($q_{1k}$).
        \end{tablenotes}
    \end{threeparttable}
    \end{table*}

\begin{table*}
    \caption{Final coefficients of $Z_{5}(G, \nu_{eff},\beta)$ obtained by joining the results in Table \ref{table:quasar_fit} and Table \ref{table:binaries_fit}.} 
    \label{table:final_fit}      
    \centering
    \begin{threeparttable}
         \begin{tabular}{c c c c c c c c c}     
            \hline\hline                 
           $G$ & $q_{00}$ & $q_{10}$ & $q_{20}$ & $q_{10}$ & $q_{11}$ & $q_{12}$ & $q_{20}$\\ 
           \hline
                       6.0 & -17.82  & +18.56 & -31.07 & -53.63 & +149.36 & -298.61 & -4877.35   \\
           
           10.8 & -13.26  & +13.52 & -30.77 & +20.86 & -10.82 & +53.55 & -2412.08   \\
           
           11.2 & -23.51  & +9.61 & -32.61 & -73.44 & +88.29 & -14.61 & +13501.35 \\
           
           11.8 & -31.06  & +16.19 & -27.02 & -44.66 & -27.31 & +44.23 & -4053.40 \\
           
           12.2 & -9.10  & +13.07 & -30.86 & -95.25 & +18.22 & +97.42 & +5590.58   \\
    
           12.9 & -11.70  & +17.77 & -34.43 & -38.31 & -83.65 & +16.80 & +56.79   \\
    
           13.1 & -30.48  & +15.50 & -31.92 & -7.78 & -67.85 & +16.60 & -272.16   \\
    
           15.9 & -33.38  & +6.62 & -20.93 & -82.04 & -48.82 & +43.17 & -2002.14   \\
    
           16.1 & -28.46 & +16.35  & -39.53 & +47.80  & -159.86 & +136.25 & -  \\
           
           17.5 & -20.79  & -6.36  & -2.21 & -55.98 & +41.07 & -20.35 & -  \\
           
           19.0 & -11.49  & +4.50  & -1.85 & -45.32  & -26.16 & +5.02 & -  \\
           
           20.0 & -5.03  & -0.69  & -5.61 & -57.69  & +49.24 & +76.35 & - \\
           
           21.0 & -2.29  & -2.01  & +51.04 & -112.23  & +123.50 & -261.33 & - \\
          \hline                                   
        \end{tabular}
        \begin{tablenotes}
        \footnotesize
            \item Notes. The table gives $q_{jk}(G)$ at the values of $G$ in the first column. For other values of $G$, linear interpolation should be used. A dash (–) indicates that the coefficient is zero. Units are: $\mu as$ (for $q_{0k}$) and $\mu as$ $\mu m$ ($q_{1k}$). 
        \end{tablenotes}
    \end{threeparttable}
     
    \end{table*}

\subsection{Combined fit\label{sec4.3}}

The magnitude intervals overlap between Table \ref{table:quasar_fit} and \ref{table:binaries_fit} for $G$ = 16.1 to 19.0. 
We combine the coefficients for $G$ = 16.1 to 21.0 from Table \ref{table:quasar_fit}  and the coefficients for $G$ = 6.0 to 15.9 from Table \ref{table:binaries_fit}. After testing, this combination demonstrates effective corrections. The end result is shown in Table \ref{table:final_fit}, which is our final estimate of the bias function for five-parameter solutions applying to the Galactic plane.

\section{Comparison and verification \label{sec5}}
In this section, we apply the new bias correction in this work (TW) and the L21 correction to various samples to evaluate and compare their effectiveness. We use quasars, used for deriving the correction in Sect \ref{sec4.1}, to perform a consistency check on the procedures used to derive the correction. Additionally, we test two corrections on globular clusters and sources with independent parallaxes from other surveys or methods, which were independent of the derivation process.

    \subsection{Using quasars \label{sec5.1}}
\begin{figure}
      \centering
      \subfigure{\includegraphics[width=0.85\linewidth]{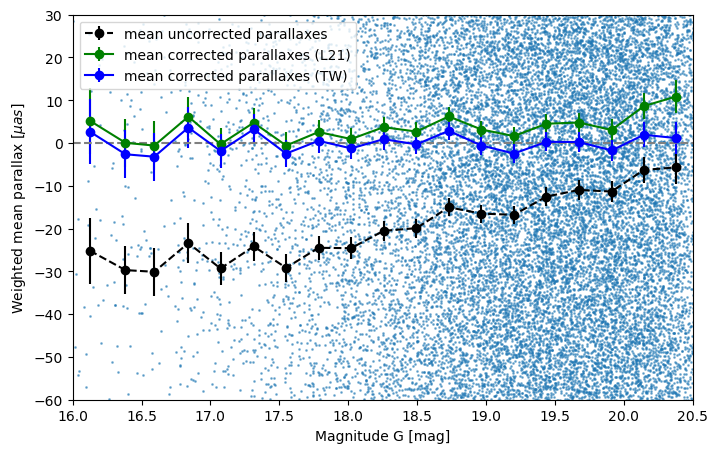}}
      \subfigure{\includegraphics[width=0.85\linewidth]{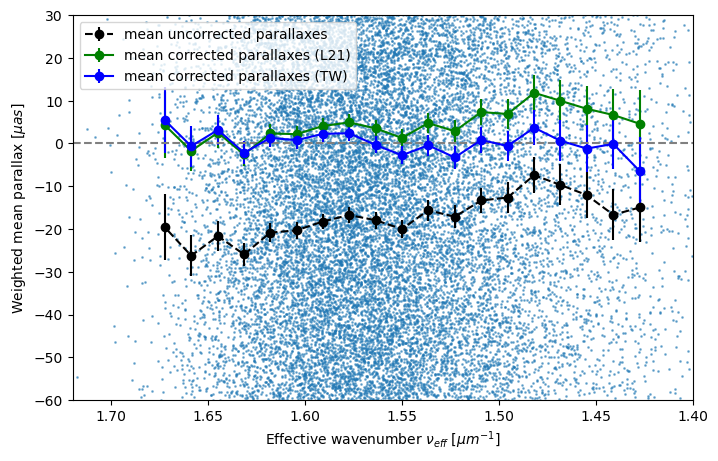}}
      \subfigure{\includegraphics[width=0.85\linewidth]{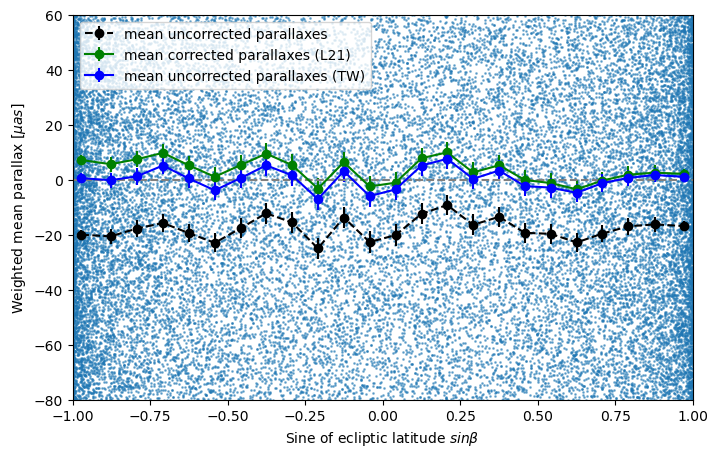}}
      \caption{ Parallaxes for 180 231 quasars with five-parameter solutions in GDR3. }
      \label{fig:valid_qso}
    \end{figure}

    We apply two bias corrections to quasars described in Sect \ref{sec4.1}, expecting the mean corrected parallaxes applying the new bias correction to be closer to zero, independent of magnitude, colour, etc. Figure \ref{fig:valid_qso} shows the results for the five-parameter solutions, divided according to magnitude, colour, and ecliptic latitude. Black circles show mean values of the uncorrected parallaxes ($\varpi$) in bins of magnitude etc.; blue circles and green circles show mean values of the corrected parallaxes ($\varpi-Z_{5}$) applying the new correction of this work (TW) and L21 correction respectively. Overall, the parallaxes corrected with the new bias correction are closer to zero than those corrected with the L21 model, indicating that the new correction is more applicable to the Galactic plane. The correction difference between L21 and this work is approximately 9.71 $\mu as$ at the faint magnitude end and go up to 11.07 $\mu as$ when $\nu_{eff}$ = 1.42 $\mu m^{-1}$.


    \subsection{Using globular clusters \label{sec5.2}}
    Since all stars in a given globular cluster share the same true parallax (with negligible spread), we can use stars in globular clusters to validate bias corrections. 
    \cite{2021MNRAS.505.5978V} utilised GEDR3 data to investigate the kinematic properties of Milky Way globular clusters, employing a mixture modelling approach to simultaneously determine cluster membership probabilities for each star and infer cluster properties, particularly the mean parallax.
    Several clusters located at $\left | b\right | \leqslant {20}^{\circ}$ are considered for our purpose, including NGC 6397, NGC 5139, NGC 6121, and NGC 6656. These clusters are located at $b \simeq {-12}^{\circ}$, ${+15}^{\circ}$, ${+16}^{\circ}$, and ${-8}^{\circ}$, with mean parallaxes of 0.416 $\pm$ 0.010 mas, 0.193 $\pm$ 0.009 mas, 0.556 $\pm$ 0.010 mas, and 0.306 $\pm$ 0.010 mas \citep{2021MNRAS.505.5978V}, respectively.
    We select stars in these clusters with five-parameter solutions that have high membership probability and reliable astrometric measurements, as described in Eq. (\ref{eq4}).
    
        \begin{align} \label{eq4}
        \left\{     
        	\begin{aligned}
                (i) \quad &qflag = 3,  \\
                (ii) \quad &memberprob >= 0.9
        	\end{aligned}
        \right.
        \end{align}
    Criterion (i) selects reliable astrometric sources  with 
    five-parameter solutions; Criterion (ii) selects stars with high membership probability (>=90\%).  Using these `clean' subsets, we compare the validity of the two bias corrections.
    
\begin{figure}
      \centering
      \subfigure{\includegraphics[width=0.85\linewidth]{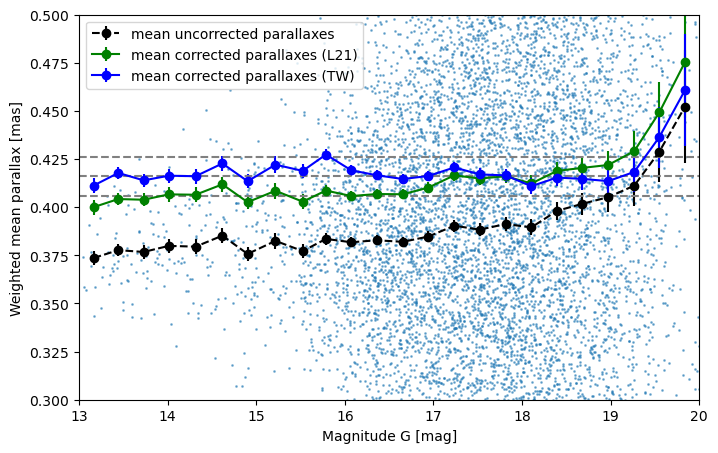}}
      \subfigure{\includegraphics[width=0.85\linewidth]{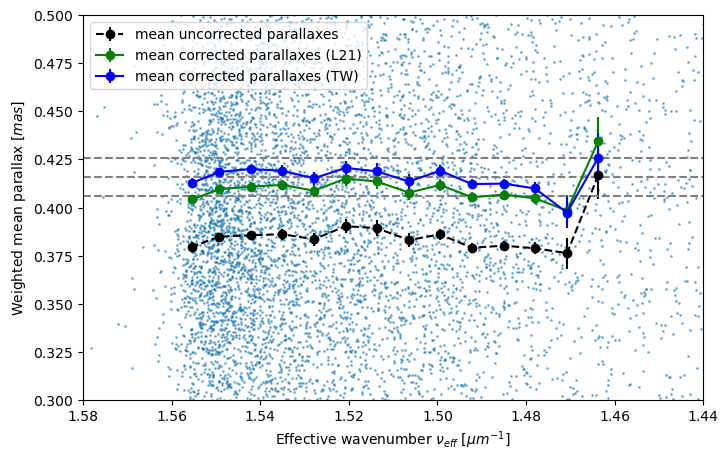}}
      \subfigure{\includegraphics[width=0.85\linewidth]{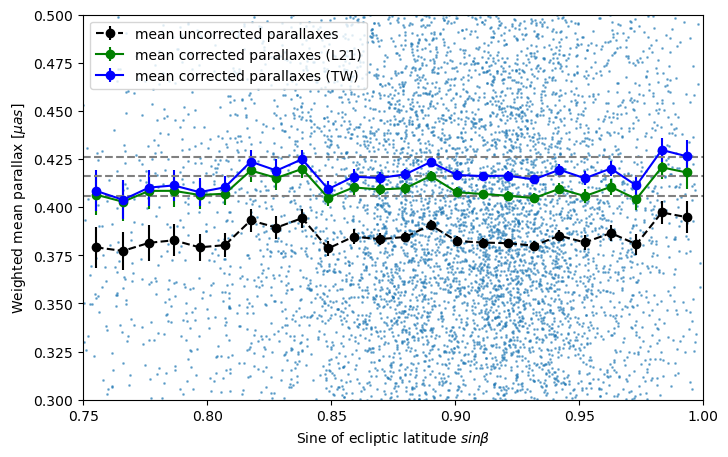}}
      \caption{ Parallaxes for 12 307 stars in NGC 6397 with five-parameter solutions in GDR3. }
      \label{fig:valid_ngc6397}
    \end{figure}

    Figure \ref{fig:valid_ngc6397} shows the results for stars in NGC 6397, divided according to magnitude, effective wavenumber and ecliptic latitude. Blue dots show the individual values plotted versus magnitude, effective wavenumber, and sine of ecliptic latitude. Black circles show mean values of the uncorrected parallaxes ($\varpi$) in bins of magnitude etc.; blue circles and green circles show mean values of the corrected parallaxes ($\varpi-Z_{5}$) applying the new correction of this work (TW) and the L21 correction respectively. The middle grey dashed line indicates the mean parallax (0.416 mas), while the other two grey dashed lines represent its statistical uncertainty (0.010 mas). Overall, the corrected parallaxes applying the new correction are closer to the mean parallax. 
    The correction difference of NGC 6397 between L21 and this work is about 10.0 $\mu as$ when 0.9 < $\sin \beta$ < 0.95, and can go up to 17.82 $\mu as$ when G = 16.
    Results for the other clusters are described in Sect. \ref{sec5.4}.
     
    \begin{table*}
    \caption{Parallax corrections for sources with independent parallaxes located at $\left | b\right | \leqslant {20}^{\circ}$ with five-parameter solutions.}           
    \label{table:valid_star}      
    \centering
    \begin{threeparttable}
         \begin{tabular}{c c c c c}     
            \hline\hline                 
           Catalogue & \makecell{$\Delta \varpi$ (uncorrected parallax offset)\\ ($\mu as$) } & \makecell{L21 correction \\ ($\mu as$) } & \makecell{TW correction \\ ($\mu as$) } & N \\ 
           \hline
           VLBI & -8.22 & +21.48  & +9.37 & 32   \\
           
           HST & -22.96  & +1.43 & +1.22 & 7  \\
           
           Red Giant  & -19.15  & +2.21  & +2.30 & 6300 \\
              
          \hline                                   
        \end{tabular}
        \begin{tablenotes}[flushleft]
        \footnotesize
            \item Notes. Column 1 gives the Name of catalogues. Column 2 gives the weighted mean of the offset between the observed GDR3 parallaxes and the independent parallaxes. Column 3 gives the weighted mean of the L21 correction. Column 4 gives the weighted mean of this work. Column 5 gives the number of sources located at $\left | b\right | \leqslant {20}^{\circ}$ with five-parameter solutions.
        \end{tablenotes} 
    \end{threeparttable}
    \end{table*}

    \subsection{Using sources with independent parallaxes from other surveys or methods\label{sec5.3}}
     
     Following the method in Sect. \ref{sec3.3}, we compare independent parallaxes from other surveys or methods with the GDR3 parallaxes to assess the validity of the two corrections in the Galactic plane. VLBI, HST and RG sources with independent parallaxes, located at $\left | b\right | \leqslant {20}^{\circ}$, are listed in Table \ref{catalogs}. Since these sources may have different external systematics, their tests serve as secondary verification.
     The results are given in Table \ref{table:valid_star}. 
     The new correction in this work (TW) proves more effective than the L21 correction for VLBI sources. Other sources show comparable results for both two corrections. Overall, our correction demonstrates effectiveness for most sources with independent parallaxes.

\subsection{Limitations \label{sec5.4}}

Results for NGC 5139, NGC 6121, and NGC 6656 can be found in Appendix \ref{appendix:b}.
Overall, the two corrections exhibit similar efficacy for these clusters, though the two corrections are not optimal. This may be due to challenges in determining globular cluster membership of \cite{2021MNRAS.505.5978V}. We acknowledge the possibility that our correction method may be sample-dependent. However, even with a more comprehensive sample for fitting the new model, sample dependence issues may persist \citep{2021A&A...654A..20G}.

\section{Summary\label{sec6}}
L21 proposed a complex recipe to address parallax bias in GDR3, drawing from quasars in GCRF3, stars in LMC, and physical binaries. However, since the L21 sample is sparse in the Galactic plane, we evaluated the efficiency of the L21 correction in this region using quasars, wide binaries, and sources with parallaxes from other surveys or methods. Results, detailed in Fig. \ref{fig:test_l21_qso_l20} - \ref{fig:test_l21_wb_g20} and Table \ref{table:test_l21_star}, suggest that the L21 correction is sample-dependent and not applicable to the Galactic plane.
We fit the parallax bias function for sources with five-parameter solutions and get the new coefficients applying to the Galactic plane, primarily based on parameterised function $Z(G, \nu_{eff},\beta)$ defined in L21 Appendix A. 
Direct estimation of the parallax bias via quasars is supplemented by indirect methods involving binaries. For sources in this region with five-parameter solutions in GDR3, the new fitted bias function is given by the functions $q_{jk}(G)$ obtained through linear interpolation in Table \ref{table:final_fit}.


In this study, we present a new parallax bias correction tailored specifically for the Galactic plane, offering improvements over the existing L21 correction where the validity for the Galactic plane was untested. The correction difference between L21 and this work can go up to 10 $\mu as$ within certain ranges of magnitude and colour. The differences in the results of the two corrections are mainly reflected in magnitude and colour, which may be caused by the influence of extinction on the Milky Way disk on the magnitude and colour. Historically, the Galactic plane has posed challenges for quasar surveys due to its high source density and significant extinctions. 
Our work represents a comprehensive effort to compile the most exhaustive collection of quasars within this region to date.
It is important to note that our paper does not claim to provide a definitive solution, but rather offers a new approach to address the parallax bias issue in the Galactic plane. This work provides an additional recipe for users of \textit{Gaia} parallaxes, especially for sources located near the Galactic plane. We acknowledge the possibility that our correction method may be sample-dependent. However, we have endeavoured to incorporate data from diverse sources within the Galactic plane and rigorously tested our correction against various samples.
Further progress is expected with \textit{Gaia} DR4, which will have improved parallax uncertainties and reduced systematics.

\begin{acknowledgements}
      We appreciate the insightful advice provided by Dr. G.A. Brown. This work has made use of data from the European Space Agency (ESA) mission \textit{Gaia} (\href{https://www. cosmos.esa.int/gaia}{https://www. cosmos.esa.int/gaia}), processed by the \textit{Gaia} Data Processing and Analysis Consortium (DPAC, \href{https://www.cosmos.esa. int/web/gaia/dpac/consortium}{https://www.cosmos.esa. int/web/gaia/dpac/consortium}). Funding for the DPAC has been provided by national institutions, in particular the institutions participating in the \textit{Gaia} Multilateral Agreement. This research has made use of the VizieR catalogue access tool and the cross-match service provided by CDS, Strasbourg. We are also very grateful to the developers of the TOPCAT \citep{2005ASPC..347...29T} software. This work has been supported by the National Natural Science Foundation of China (NSFC) through grants 12173069, the Youth Innovation Promotion Association CAS with Certificate Number 2022259, the Talent Plan of Shanghai Branch, Chinese Academy of Sciences with No.CASSHB-QNPD-2023-016, the Strategic Priority Research Program of the Chinese Academy of Sciences, Grant No.XDA0350205. We acknowledge the science research grants from the China Manned Space Project with NO. CMS-CSST-2021-A12 and NO.CMS-CSST-2021-B10.
\end{acknowledgements}

\bibliographystyle{aa}
\bibliography{myref}

\begin{appendix}

\section{ADQL queries for the galaxy and star sample\label{appendix:a}}

Figure \ref{fig:quasar_color_color} shows the distribution of the two colours of Non-GCRF3 quasars, GCRF3 quasars, galaxies, and stars located at $\left | b\right | \leqslant {20}^{\circ}$. The sources used to construct samples of galaxies and stars are as follows:
\begin{itemize}
    \item Galaxies: random 100 000 galaxies located at $\left | b\right | \leqslant {20}^{\circ}$ using criteria provided in \citet[see Table~11]{2023A&A...674A..41G} to select the pure galaxy sample. The ADQL query is as follows. 

\begin{verbatim}
SELECT TOP 100000 g.*
FROM gaiadr3.galaxy_candidates as c
INNER JOIN gaiadr3.gaia_source as g
USING (source_id)
WHERE (c.radius_sersic IS NOT NULL
OR c.classlabel_dsc_joint='galaxy' 
OR c.vari_best_class_name='GALAXY') 
AND (g.b BETWEEN -20 AND 20)
\end{verbatim}

    \item Stars: We take a random 100 000 stars located at $\left | b\right | \leqslant {20}^{\circ}$ by selecting high reliable astrometric sources which are not candidates of quasars and galaxies. The ADQL query is as follows. 

\begin{verbatim}
SELECT TOP 100000 *
FROM gaiadr3.gaia_source
WHERE in_qso_candidates = 'false' 
AND in_galaxy_candidates = 'false'
AND ruwe <= 1.4 
AND parallax <= 0.2
AND parallax_over_error >= 5 
AND astrometric_excess_noise_sig <= 2 
AND astrometric_excess_noise <= 0.1
AND visibility_periods_used >11
AND duplicated_source = 'false'
AND phot_variable_flag !='VARIABLE'
AND phot_bp_mean_flux_over_error > 100
AND phot_g_mean_flux_over_error > 100
AND phot_rp_mean_flux_over_error > 100
AND b BETWEEN -20 AND 20
\end{verbatim}

\end{itemize}

\section{Parallax differences of binaries against magnitude and colour differences}\label{appendix:parallax diff}

\begin{figure}
              \centering
              \subfigure{\includegraphics[width=0.85\linewidth]{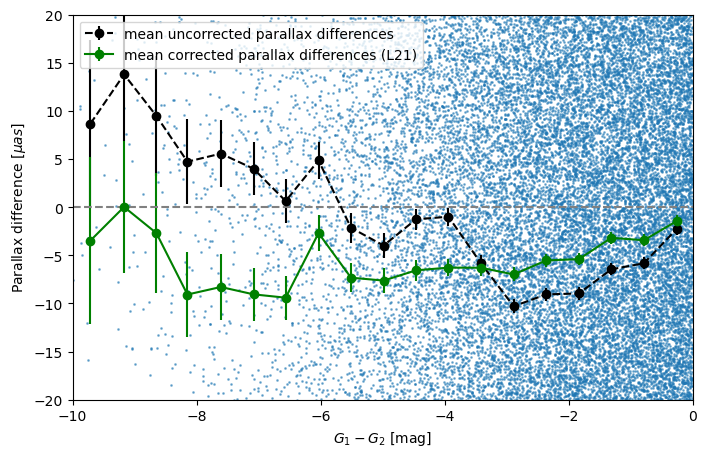}}
              \subfigure{\includegraphics[width=0.85\linewidth]{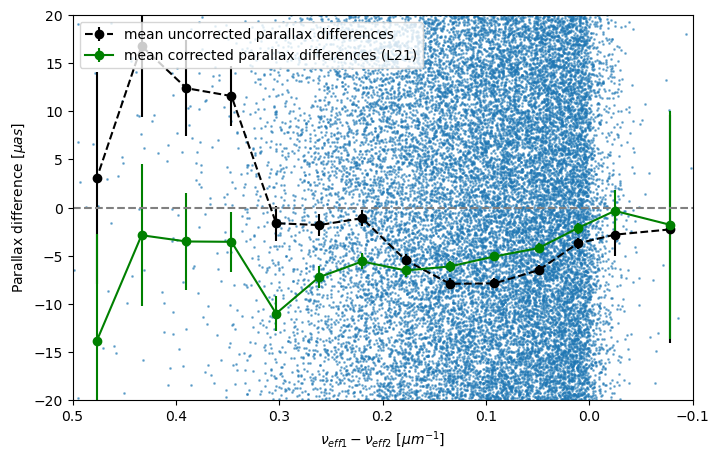}}
              \caption{Parallax differences of 148 234 binaries located at $\left | b\right | \leqslant {20}^{\circ}$ with five-parameter solutions versus magnitude and colour differences. }
              \label{fig:test_l21_wb_l20_add}
        \end{figure}

\begin{figure}
              \centering
              \subfigure{\includegraphics[width=0.9\linewidth]{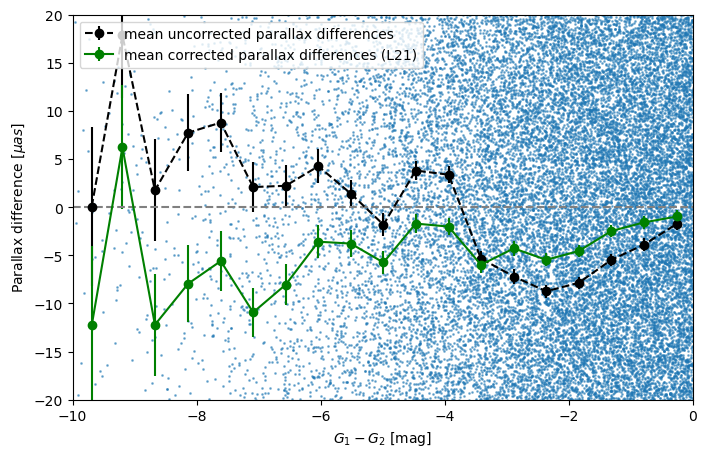}}
              \subfigure{\includegraphics[width=0.9\linewidth]{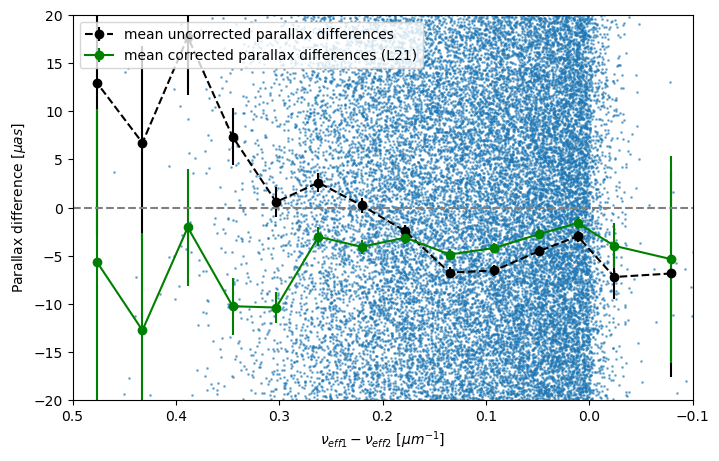}}
              \caption{Parallax differences of 172 931 binaries located at $\left | b\right | > {20}^{\circ}$ with five-parameter solutions versus magnitude and colour differences.}
              \label{fig:test_l21_wb_g20_add}
        \end{figure}

Figure \ref{fig:test_l21_wb_l20_add} and \ref{fig:test_l21_wb_g20_add} show the results for the five-parameter solutions of WBs located at $\left | b\right | \leqslant {20}^{\circ}$ and $\left | b\right | > {20}^{\circ}$ according to magnitude difference ($G_{1}-G_{2}$) and colour difference ($\nu_{eff1}-\nu_{eff2}$), respectively. 
The parallax differences go to zero for small magnitude and colour differences for both the corrected and uncorrected versions.
        
\section{Parallax zero-point differences of sources with derived parallaxes against colour and ecliptic latitude}\label{appendix:star_diff}

\begin{figure*}
              \centering
              \subfigure[$\left | b\right | \leqslant {20}^{\circ}$]
              {
                \begin{minipage}[b]{.48\linewidth}
                \centering
                \includegraphics[width=1\linewidth]{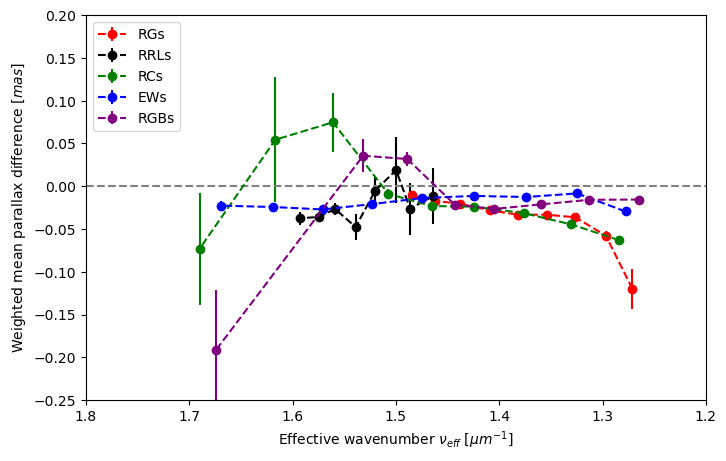} \\
                 \includegraphics[width=1\linewidth]{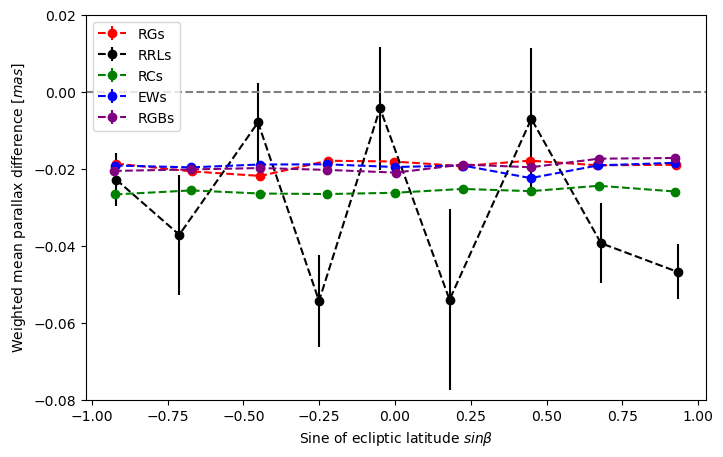}
         
                \end{minipage}
              }
              \subfigure[$\left | b\right | > {20}^{\circ}$]
              {
              \begin{minipage}[b]{.48\linewidth}
               \centering
              \includegraphics[width=1\linewidth]{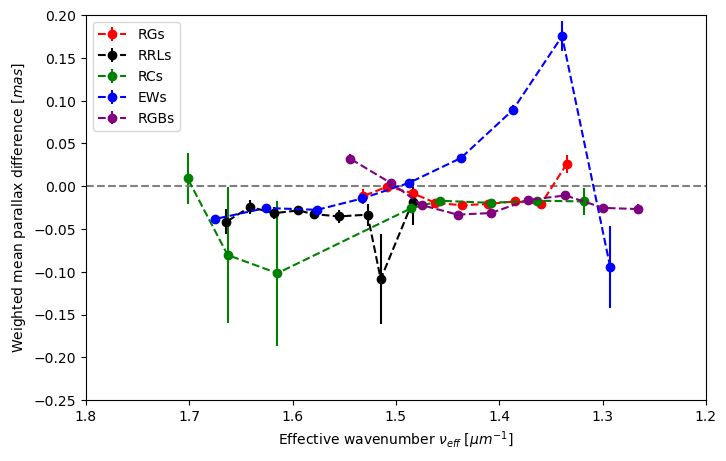}\\
              \includegraphics[width=1\linewidth]{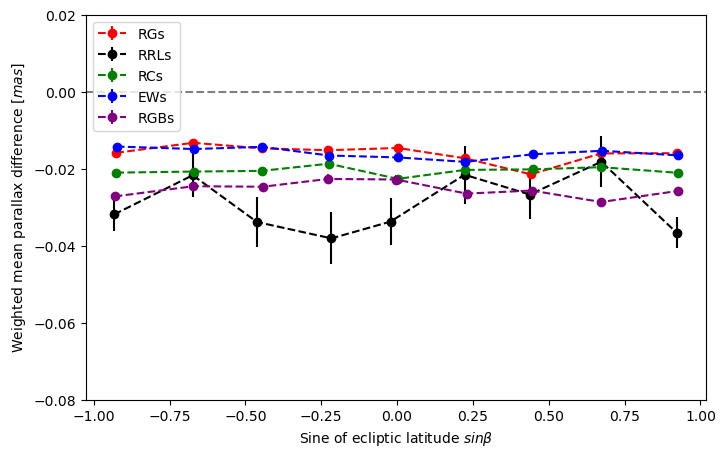}
              \end{minipage}
             
              }
            \caption{Parallax zero-point differences versus colour and ecliptic latitude of sources with derived parallaxes (RGs, RRLs, RCs, EWs, and RGBs) located at $\left | b\right | \leqslant {20}^{\circ}$ (a) and  $\left | b\right | > {20}^{\circ}$ (b) with five-parameter solutions. The dots with error bars represent the weighted mean parallax differences for each bin across different samples. The large error bars for RRLs (black) are attributed to the limited sample size. The grey dashed line indicates a parallax difference of zero.
              }
            \label{fig:valid_star}
        \end{figure*}

        Figure \ref{fig:valid_star} shows the results for the five-parameter solutions of sources with derived parallaxes from different methods (RGs, RRLs, RCs, EWs, and RGBs), located at $\left | b\right | \leqslant {20}^{\circ}$ (a) and $\left | b\right | > {20}^{\circ}$ (b) according to colour and ecliptic latitude, respectively. VLBI and HST sources are not plotted due to their small sample sizes. Overall, the parallax differences show clear distinctions between regions within and outside the Galactic plane.

\section{Tests of other globular clusters }\label{appendix:b}


    \begin{figure}
      \centering
      \subfigure{\includegraphics[width=1.0\linewidth]{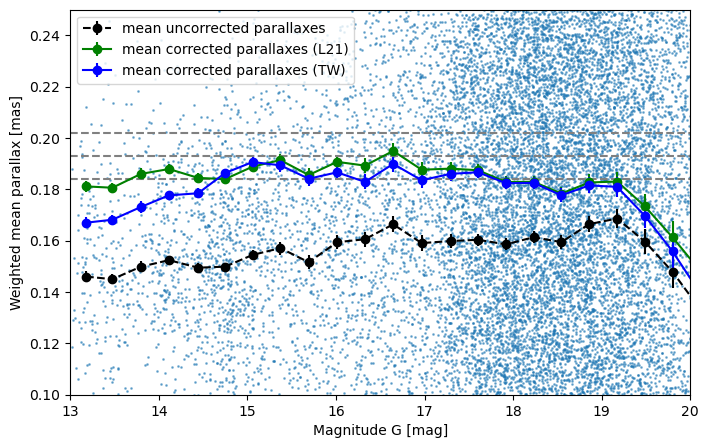}}
      \subfigure{\includegraphics[width=1.0\linewidth]{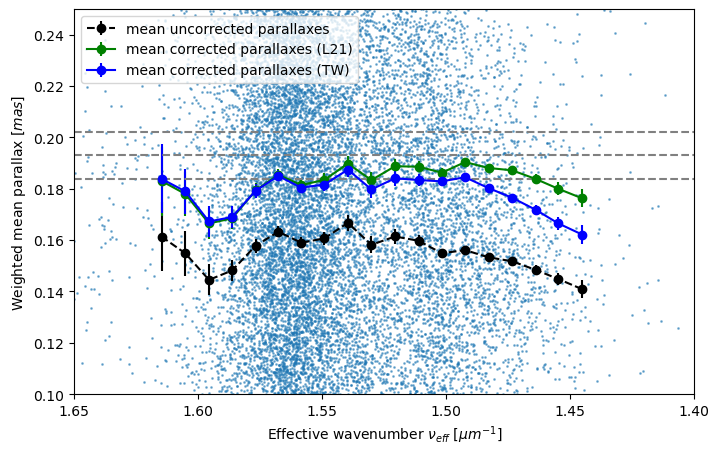}}
      \subfigure{\includegraphics[width=1.0\linewidth]{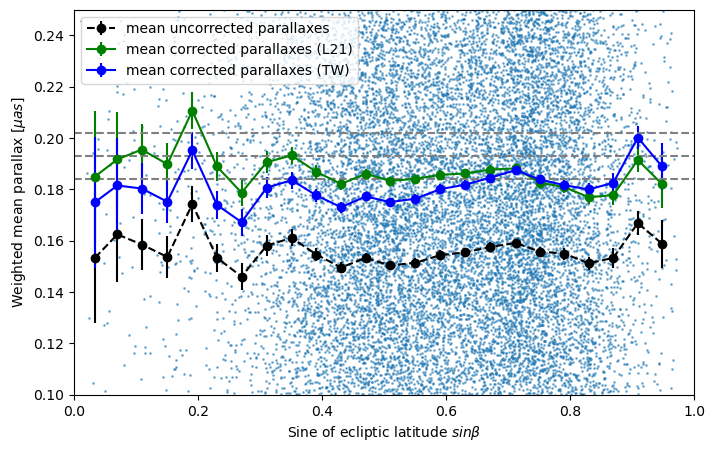}}
      \caption{ Parallaxes for 53 309 stars in NGC 5139 with five-parameter solutions in GDR3. The middle grey dashed line indicates the mean parallax (0.193 mas), while the other two grey dashed lines represent its statistical uncertainty (0.009 mas).}
      \label{fig:valid_ngc5139}
    \end{figure}

      \begin{figure}
      \centering
      \subfigure{\includegraphics[width=1.0\linewidth]{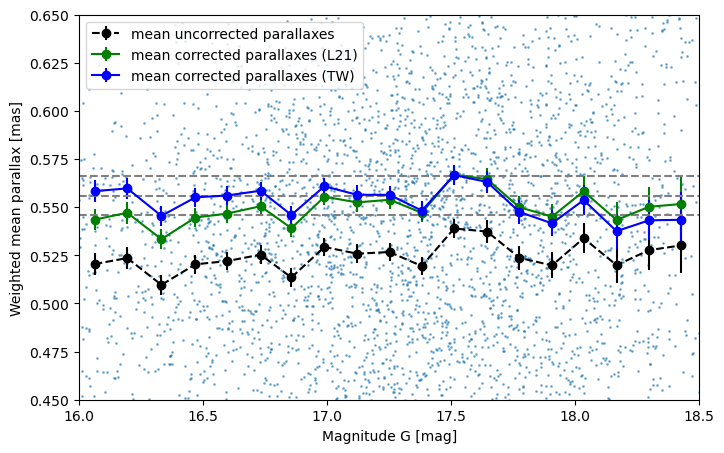}}
      \subfigure{\includegraphics[width=1.0\linewidth]{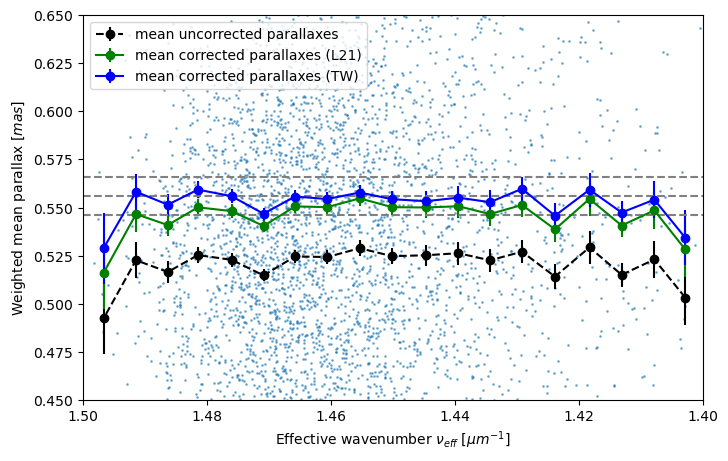}}
      \subfigure{\includegraphics[width=1.0\linewidth]{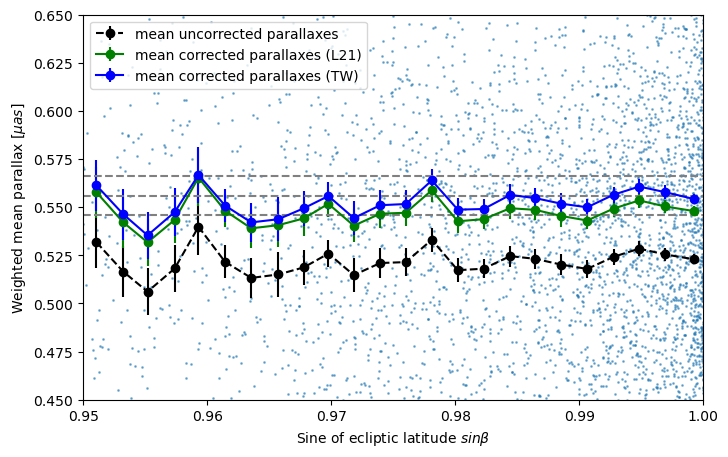}}
      \caption{ Parallaxes for 5208 stars in NGC 6121 with five-parameter solutions in GDR3. The middle grey dashed line indicates the mean parallax (0.556 mas), while the other two grey dashed lines represent its statistical uncertainty (0.010 mas).}
      \label{fig:valid_ngc6121}
    \end{figure}

      \begin{figure}
      \centering
      \subfigure{\includegraphics[width=1.0\linewidth]{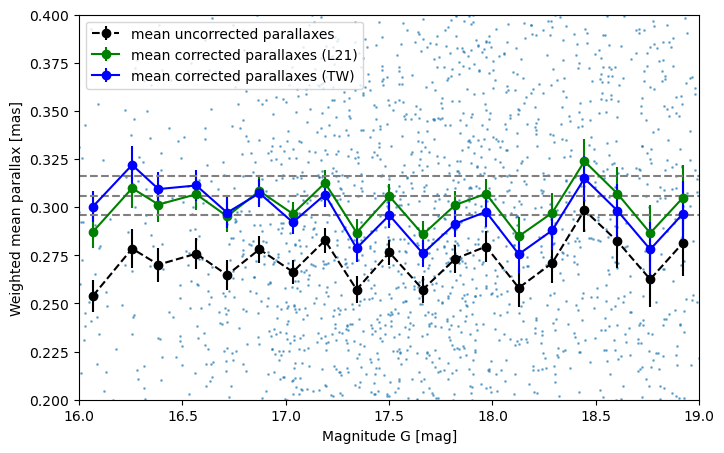}}
      \subfigure{\includegraphics[width=1.0\linewidth]{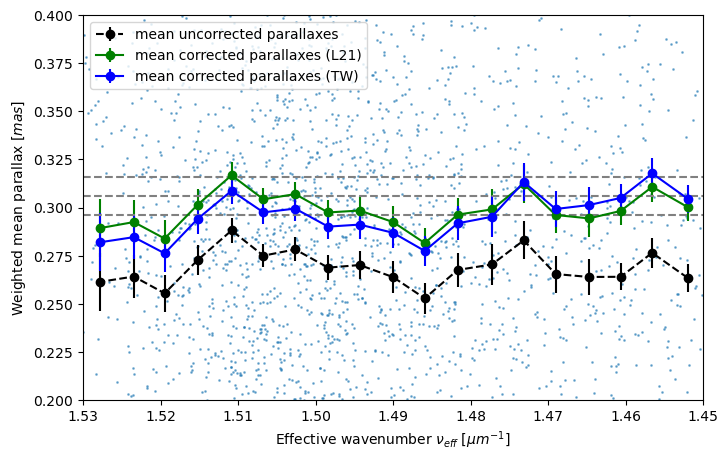}}
      \subfigure{\includegraphics[width=1.0\linewidth]{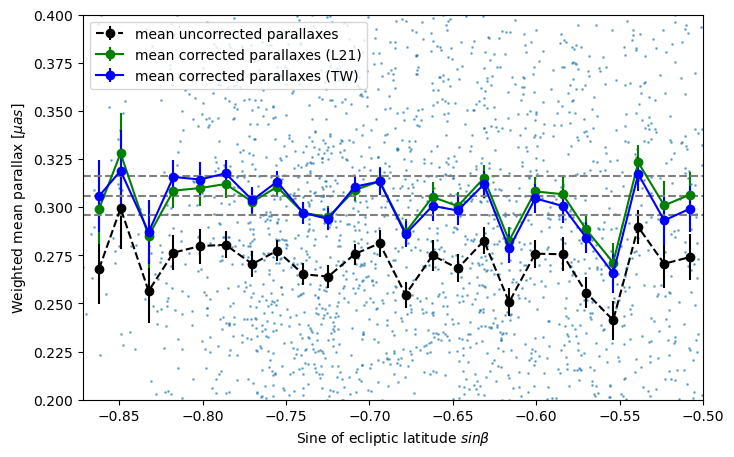}}
      \caption{ Parallaxes for 4323 stars in NGC 6656 (0.306 $\pm$ 0.010 mas) with five-parameter solutions in GDR3. The middle grey dashed line indicates the mean parallax (0.306 mas), while the other two grey dashed lines represent its statistical uncertainty (0.010 mas).}
      \label{fig:valid_ngc6656}
    \end{figure}

    Figure \ref{fig:valid_ngc5139}, \ref{fig:valid_ngc6121}, and \ref{fig:valid_ngc6656} show the results for stars in NGC 5139, NGC 6121, and NGC 6397, respectively. These figures are divided according to magnitude, effective wavenumber and ecliptic latitude. Blue dots show the individual values plotted versus magnitude, effective wavenumber, and sine of ecliptic latitude. Black circles show mean values of the uncorrected parallaxes ($\varpi$) in bins of magnitude etc.; blue circles and green circles show mean values of the corrected parallaxes ($\varpi-Z_{5}$) applying the new correction in this work (TW) and the L21 correction respectively. 
    In Fig. \ref{fig:valid_ngc5139}, the corrected parallaxes applying the L21 model (green) of NGC 5139 are closer to the mean parallax zone (0.193 $\pm$ 0.009 mas). Figures \ref{fig:valid_ngc6121} and \ref{fig:valid_ngc6656} show that both corrections yield comparable results within the well-populated region. 

\end{appendix}

\end{document}